\documentclass[11pt,a4paper]{article}
\usepackage[utf8]{inputenc}

\usepackage[a4paper, margin=1in]{geometry}

\usepackage{bm}
\usepackage{graphicx}
\usepackage{newtxtext}
\usepackage{newtxmath}
\usepackage{natbib}
\usepackage{hyperref}
\hypersetup{
    colorlinks = true,
    urlcolor   = blue,
    citecolor  = black,
}

\newcommand{\RomanNumeralCaps}[1]
\linenumbers

\usepackage[nameinlink]{cleveref}
\crefname{equation}{}{}
\crefname{figure}{figure}{figures}
\usepackage{subcaption,color}
\usepackage[singlelinecheck=false]{subcaption}
\usepackage{placeins}
\usepackage{overpic}

\title{Assessing passive scalar dynamics in bubble-induced turbulence using DNS}

\author{Niklas Hidman$^1$,
  Henrik Str\"om$^1$,
  Srdjan Sasic$^1$ and Gaetano Sardina$^1$}

\date{%
    $^1$Department of Mechanical and Maritime Sciences, Chalmers University of Technology, Gothenburg, Sweden\\[2ex]%
    \today
}

\begin{document}

\maketitle

\begin{abstract}
By using Direct Numerical Simulations (DNS) of bubbly flows with passive scalars, we show a transition in the scalar spectra from a $k^{-5/3}$ to a $k^{-3}$ scaling with the wavenumber $k$, in contrast with those of single-phase isotropic turbulence. For cases with a mean scalar gradient in the horizontal direction, the scalar spectrum decays faster than $k^{-3}$ at high wavenumbers. While the $k^{-3}$ scaling is well established in the bubbly flow velocity spectrum, the scalar spectrum behaviour is not fully understood. We find that the transition length scale of the scalar spectra is comparable to or below the bubble diameter and decreases with the molecular diffusivity of the scalar in the liquid. We use DNS to compute the scalar spectra budget and show that the scalar fluctuations are produced by the mean scalar gradient at length scales above the bubble diameter, contrary to the velocity fluctuations. At length scales below the bubble diameter, the net scalar transfer scales as $k^{-1}$ inducing the $k^{-3}$ scaling of the scalar spectra. This finding is consistent with the hypothesis proposed by \cite{dung2022emergence} about the physical mechanism behind the $k^{-3}$ scaling. We also show dependencies of the bubble suspension's convective scalar diffusivity on the gas volume fraction and  molecular diffusivity that differs based on the direction of the mean scalar gradient. For a mean scalar gradient in the vertical direction, we find and qualitatively explain a significant effect of the molecular diffusivity in the gas on the convective scalar diffusivity.
\end{abstract}

\section{Introduction}

The transport of scalar fields, such as heat or chemical species, in bubbly flows occurs in many industrial and natural processes such as chemical reactors, heat exchangers and atmosphere-ocean exchanges. Empirical evidence suggests that bubbly flows enhance the transport of scalars without moving parts, diminishing costs \citep{Mudde2005}.

Bubbles rising in a liquid at moderate volume fractions induce peculiar velocity fluctuations commonly known as bubble-induced turbulence \citep{lance_bataille_1991, Risso2018}. Experiments  \citep{mercado2010, Riboux2010, Mendez-Diaz2013} and numerical studies \citep{Pandey2020, Innocenti2021} have shown a robust power-law scaling with an exponent of $-3$ in the kinetic energy spectrum at an approximate interval of wavenumbers $k$ below the bubble diameter before viscous dissipation occurs. This scaling differs from the classical $-5/3$ observed in single-phase homogeneous isotropic turbulence.

Although the properties of bubble-induced turbulence have been thoroughly investigated, the dynamics and statistics of a passive scalar in such flows have only recently received attention. For example, \cite{Almeras2015} examined experimentally the mixing of a low-diffusive dye in homogeneous bubbly flows and showed that the scalar dispersion could be modelled as an anisotropic diffusion process with the effective diffusivity $\propto \phi^{0.4}$ at low gas volume fractions $\phi$. A follow-up work \citep{Almeras2016} shows, in the same experimental configuration, a -3 scaling of the scalar spectrum in the frequency domain.
\cite{Loisy2018} studied numerically the passive scalar mixing in bubbly flows of up to $12$ bubbles at low bubble Reynolds numbers ($Re=30$). That study showed that the convective contribution (due to bubble-induced agitation) to the effective scalar diffusivity is dominant for most common bubbly flows.
\cite{Gvozdic2018} experimentally investigated the heat transport in a bubble column heated on one lateral side and cooled on the other. They found the effective thermal diffusivity in the horizontal direction $\propto \phi^{0.45}$ for $\phi\leq5\%$. They also observed a scaling of $-1.4$ in the temperature spectrum at frequencies around $0.1-3$ $Hz$; however, the used thermistors could not resolve the higher frequencies present in the bubble-induced turbulence.  
\cite{dung2022emergence} studied experimentally the thermal spectra scaling of a thermal mixing layer in vertical channel bubbly flow with an active turbulent grid. They showed a spectrum scaling transition from $-5/3$ to a $-3$ scaling in the frequency domain for large enough $\phi$ that clearly shows the bubbles influence the thermal spectra. The work provides hypotheses and scaling arguments for the existence of the -3 scaling in the scalar spectrum.

For the first time, we investigate, using multiphase Direct Numerical Simulations (DNS), the spectral scaling of a passive scalar $c$ in bubble-induced turbulence. Specifically, we study the influence of the liquid Schmidt number and $\phi$ on the scalar spectra (at length scales of $O(10)$ bubble diameters down to the viscous dissipation scales) and the effective scalar diffusivity. Statistically-steady scalar fluctuations are generated by imposing constant scalar gradients in both the vertical and horizontal directions. Furthermore, we compute the energy budget of the scalar spectra to assess the hypothesis proposed in \cite{dung2022emergence} that the spectral scalar transfer $T_c(k) \propto k^{-1}$ where the scalar spectra show a $-3$ scaling. In addition, we compare the results from our bubbly flow simulations with those obtained in single-phase isotropic turbulence to elucidate the effects of the bubbles on the scalar dynamics.

\section{Methodology} \label{Sec:method}

\subsection{Problem Statement}
We numerically study the statistics of a passive scalar field with an imposed constant gradient in a fully periodic homogeneous bubbly flow domain where the bubbles agitate an initially quiescent liquid. We assume the scalar field continuous across the bubble interfaces and the surface tension, densities, viscosities and scalar diffusivities constant in the two phases. All variables are made non-dimensional using the spherical equivalent bubble diameter $d_b$, characteristic rise velocity $\sqrt{g d_b}$ ($g$ is the gravitational acceleration) and the liquid density $\rho_l$.
The problem is completely described by the seven, a priori known, dimensionless parameters; the E\"otv\"os number $\mathit{Eo} = {\rho_l g d_b^2}/{\sigma}$ relating buoyancy to surface tension forces, the Galilei number $\mathit{Ga} = {\rho_l \sqrt{gd_b}d_b}/{\mu_l}$ that is the ratio of buoyancy to viscous forces, the density ratio $\rho_r={\rho_g}/{\rho_l}$, the dynamic viscosity ratio $\mu_r = {\mu_g}/{\mu_l}$, the gas volume fraction $\phi = N_b\pi d_b^3/(6L^3)$ and the Schmidt numbers $\mathit{Sc}_l = \nu_l/D_l$ and $\mathit{Sc}_g = \nu_g/D_g$. Here, $\sigma$ is the surface tension, $N_b$ the number of bubbles, $L$ the side length of the cubic domain, $\nu=\mu/\rho$ the kinematic viscosity and the subscripts $l$ and $g$ denote the liquid and gas phases. We use $N_b \geq 40$ monodisperse bubbles to get statistics independent of $N_b$ \citep{Loisy2017}. We choose $\rho_r=10^{-3}$ and $\mu_r=10^{-2}$ that resemble air-water systems. These ratios are very small and the exact values are physically insignificant for most gas-liquid systems of practical interest \citep{Bunner2002}. We fix the $\mathit{Ga}=390$ and $\mathit{Eo}=0.85$ that correspond to $2.5$ $mm$ air bubbles in water. These parameters are characteristic of practically relevant systems and several experimental studies \citep{Riboux2010, Mendez-Diaz2013, Almeras2015, Gvozdic2018, dung2022emergence}.

We study a scalar field, such as the concentration of a chemical species or the temperature, that can be assumed a passive scalar if the effect of the scalar on the fluid properties (such as viscosity and density) is small (and the effects of viscous heating can be ignored) \citep{Loisy2018, Gvozdic2018}. Although we choose to use typical parameters in the context of mass transfer, the results are relevant for heat transport, given that the assumptions mentioned above hold.
We analyse the effects of the scalar gradient direction, liquid Schmidt numbers $\mathit{Sc}_l=(0.7, 1.5, 3, 7)$ and the gas volume fractions $\phi=1.7\%$ and $5.2\%$ on the scalar spectra and transport properties. The complete set of DNS parameters is shown in \cref{tab:DNSparams}. Note that for each case in \cref{tab:DNSparams}, we simulate two independent scalar fields with an imposed constant gradient in the vertical $\nabla^v \langle c \rangle$ and horizontal $\nabla^h \langle c \rangle$ direction, respectively. 

Additionally, we study the scalar dynamics in single-phase isotropic turbulence to compare to those in bubbly flows. We use a fully periodic domain and the same $\mathit{Sc}_l$-numbers as in the bubbly flow cases. Turbulence is maintained by an isotropic stochastic forcing localized at a wavenumber where the bubble-induced velocity spectrum shows a maximum and the dissipation is kept equal to the one of the bubble suspension $\epsilon=\phi g V_d$.

\begin{table}
\centering
    \begin{tabular}{
    p{0.8cm}p{0.8cm}p{0.8cm}p{0.8cm}p{0.8cm}p{0.8cm}p{0.8cm}p{0.8cm}p{0.8cm}p{0.8cm}p{1.cm}p{.8cm}p{.8cm}  }
     Case & $\phi \%$& $\mathit{Sc}_l$ & $\mathit{Sc}_g$ & $N_b$ & $\mathit{Re}_d$ & $u_{l,\mathit{std}}$ & $c^v_{l,\mathit{std}}$ & $c^h_{l,\mathit{std}}$ & $\epsilon^v_{l,\mathit{c}}$ & $\epsilon^h_{l,\mathit{c}}$ & $\mathit{Sh}^v$ & $\mathit{Sh}^h$\\[3pt]
     A1 & 1.7 & 0.7 & 0.7 & 40 & 717 & 0.24 & 1.69 & 0.76 &	0.25 &	0.048 & 83 & 12
\\
     A2 & 1.7 & 1.5 & 0.7 & 40 & 717 & 0.24 & 1.87 & 0.83	&0.27 & 0.051 & 161 & 23
\\
     A3 & 1.7 & 3.0 & 0.7 & 40 & 717 & 0.24 & 2.02	&0.90&	0.29	&0.052 & 270 & 38
\\
     A4 & 1.7 & 7.0 & 0.7 & 40 & 717 & 0.24 & 2.17&	0.99&	0.29&	0.053 & 422 & 59
\\
     A5 & 1.7 & 0.7 & 7.0 & 40 & 717 & 0.24 & 1.92	&0.83&	0.26&	0.051 & 115 & 14
\\
     B1 & 5.2 & 0.7 & 0.7 & 122 & 666 & 0.35 & 1.45	&0.79&	0.31	&0.067 & 99 & 13
\\
     B2 & 5.2 & 1.5 & 0.7 & 122 & 666 & 0.35 & 1.63& 0.84&	0.34&	0.069 & 168 & 20
\\
     B3 & 5.2 & 3.0 & 0.7 & 122 & 666 & 0.35 & 1.77& 0.88	&0.37&	0.070 & 236 & 28
\\
     B4 & 5.2 & 7.0 & 0.7 & 122 & 666 & 0.35 & 1.93& 0.79&	0.37&	0.063 & 296 & 34
\\
     B5 & 5.2 & 0.7 & 7.0 & 122 & 666 & 0.35 & 1.40& 0.74&	0.32	&0.065 & 194 & 19
\\
     S1 & - & 0.7 & - & - & - & 0.30 & - & 0.72 & - & 0.103 & - & 28 \\
     S2 & - & 1.5 & - & - & - & 0.30 & - & 0.80 & - & 0.107 & - & 63 \\
     S3 & - & 3.0 & - & - & - & 0.30 & - & 0.86 & - & 0.108 & - & 126 \\
     S4 & - & 7.0 & - & - & - & 0.30 & - & 0.95 & - & 0.108 & - & 297 \\
\end{tabular}
\caption{Nondimensional parameters of the DNS simulations.
$\mathit{Re}_d=V_d d_b / \nu_l$ is defined with the drift velocity $V_d$, the average relative velocity between the gas and liquid phases. $u_{l,\mathit{std}}$ and $c_{l,\mathit{std}}$ are the total standard deviation of the liquid velocity and liquid scalar fluctuations, respectively. $\epsilon_{l,\mathit{c}}=D_l\langle \nabla c'_l \cdot \nabla c'_l \rangle$ are the scalar dissipation rates in the liquid and $\mathit{Sh} = \mathsf{D}_{\mathit{conv}}/D_{\mathit{mol,s}}$ is the Sherwood number. The superscripts $v$ and $h$ represent values for the cases with an imposed scalar gradient in the vertical $\nabla^v \langle c \rangle$ and horizontal $\nabla^h \langle c \rangle$ directions, respectively. For the single-phase cases S1-4, despite the isotropy, the scalar statistics are shown in the columns for the horizontal direction.}
\label{tab:DNSparams}
\end{table}

\subsection{Numerical method}
 The scalar field is decomposed in $ c =    \nabla\langle c \rangle\cdot \boldsymbol{x}  + c' $, where $\langle c \rangle$ represents the mean scalar field that we specify as a constant slope linear field ($\nabla \langle c \rangle=1$) and $c'$ is the scalar disturbance due to the bubbles' motion that we solve numerically. The bubbly suspension is solved  using the Volume of Fluid (VOF) approach, where the two phases are tracked using a volume fraction field $f$ that is equal to 1 in the liquid and 0 in the gas. The governing equations read:
\begin{align}
     \nabla \boldsymbol{\cdot} \boldsymbol{u} &= 0\, ,\label{eq:cont} 
     \end{align}
 \begin{align}
 \rho\dfrac{D \boldsymbol{u}}{D t} &= (\rho -\langle \rho \rangle) \boldsymbol{g} - \nabla p + \dfrac{1}{Ga}\nabla \boldsymbol{\cdot}(2\mu\bm{\mathsf{S}}) + \dfrac{\kappa\delta_S \hat{\boldsymbol{n}}}{Eo}\, \label{eq:mom} ,
      \end{align}
     \begin{align}  
 \dfrac{\partial f}{\partial t} + \nabla \boldsymbol{\cdot} (\boldsymbol{u}f) &= 0\, ,
 \end{align}
      \begin{align}  
 \dfrac{\partial c'}{\partial t} + \underbrace{ \boldsymbol{u}\boldsymbol{\cdot}\nabla c'}_{h_c} &= \underbrace{\nabla \boldsymbol{\cdot} (D \nabla (\langle c \rangle + c'))}_{d_c} - \boldsymbol{u} \boldsymbol{\cdot} \nabla \langle c \rangle\, , \label{eq:scalar_transport}
\end{align}
where $\boldsymbol{u}$ is the velocity, $\boldsymbol{g}$ the gravity vector, $p$ the pressure, $\bm{\mathsf{S}}=(\nabla\boldsymbol{u} + \nabla\boldsymbol{u}^T)/2$ the rate of deformation tensor and $\kappa$ and $\hat{\boldsymbol{n}}$ are the interface curvature and normal. The additional body force $\langle \rho \rangle \boldsymbol{g}$ in \cref{eq:mom} prevents the flow from accelerating in the gravitational direction \citep{Bunner2002}. The density $\rho$ and diffusivity $D$ are the arithmetic means with $f$ of their single-phase counterparts, while $\mu$ is the harmonic mean that better approximates gas-liquid interfaces with continuous shear stress \citep{Tryggvason2011}. 

The governing equations are solved with the open-source code Basilisk \citep{Popinet2015} in a cubic periodic domain with the side length $L/d_b = 10.72$. We discretise the domain with equidistant grid points in each direction. The number of grid points is selected to resolve the smallest length scales of the velocity and scalar fields. The Kolmogorov length scale $\eta_u = (\nu^3/\epsilon)^{1/4}$ is estimated using that the dissipation rate equals the power of the buoyancy force per unit mass $\epsilon = \phi g V_d$ \citep{Risso2018}.
For  cases B, with $\phi=5.2\%$, we obtain $\eta_u / d_b=0.021$. Using the DNS resolution criterion reported in \citet{pope2001}, a grid spacing $\Delta$ satisfying $\Delta/\eta_u \lessapprox 2.1$ gives a good resolution of the smallest turbulent scales. These values indicate that a grid spacing of about $24 \Delta / d_b$ is sufficient for resolving the velocity field.

The Batchelor length scale is estimated according to $\eta_c = \eta_u / \mathit{Sc}_l^{1/2}$ \citep{Batchelor1959}. This relation indicates that for the cases B with $\mathit{Sc}_l \approx 1$, the resolution of $24 \Delta / d_b$ is sufficient, while for the cases with $\mathit{Sc}_l = 3$ and $7$, we need approximately $40$ and $60 \Delta / d_b$, respectively. The same analysis for cases A, with $\phi=1.7\%$, shows that $\mathit{Sc}_l = 3$ and $7$ require about $30$ and $50  \Delta / d_b$, respectively. 

However, the DNS resolution criterion by \citet{pope2001} is based on studies on single-phase isotropic turbulence, and, in principle, even smaller scales may be present in the bubbly flow velocity and scalar fields (such as thin boundary layers). For these reasons (and because we evolve multiple scalar fields in a single DNS), we ran all cases with $\mathit{Sc}_l = 0.7$ and $1.5$ with $48 \Delta / d_b$ and the cases with $\mathit{Sc}_l = 3$ and $7$ using another refinement level corresponding to $96 \Delta / d_b$.

The Basilisk code has been validated and used extensively for multiphase DNS of bubbly flows \citep{Innocenti2021,hidman2022}. 
The code features a finite volume solver using a time-splitting projection method with standard second-order gradient discretisation and the Bell-Colella-Glaz second-order upwind scheme for the velocity and scalar advection. We use the cell-centred Cartesian multigrid solver that efficiently solves the Helmholtz-Poisson type problems for the velocity components and the Poisson equation for the pressure correction \citep{POPINET2009}. This allows us to perform high-resolution 3D simulations with a uniform grid at a feasible computational cost. The surface tension force in \cref{eq:mom} is discretised with a well-balanced method, and the height-function approach is used to compute the interface curvature \citep{Popinet2018}. The piecewise-linear interface reconstruction technique is used to advect the volume fraction field without interface smearing \citep{Scardovelli1999}.
To avoid bubble coalescence, we implement a repulsive force by locally increasing the surface tension to $\sigma_{\mathit{rep}} = 2.1\sigma$ \citep{Talley2017} only at the part of the bubble interface that is less than $d_b$ from another bubble's centre of mass.

The single-phase simulations are performed with a pseudo-spectral solver \citep{Sardina2015}. The grid increases from 384 to 640 collocation points with increasing $Sc$ .

\section{Results} \label{Sec:results}

This section presents the passive scalar spectra and statistics from our DNS simulations of single-phase isotropic turbulence and homogeneous bubbly turbulent suspensions. The simulation cases and parameters are shown in \cref{tab:DNSparams}.

\subsection{Characteristics of the scalar dynamics} \label{sec:char_scalar_dyn}

\begin{figure}[h]
    \centering
  \begin{subfigure}[b]{0.48\linewidth}
    \caption{}

    \label{fig:c_tot_3D_vert}
    \begin{overpic}[trim = .3cm .3cm 0.3cm 0.3cm,clip,width=\linewidth]{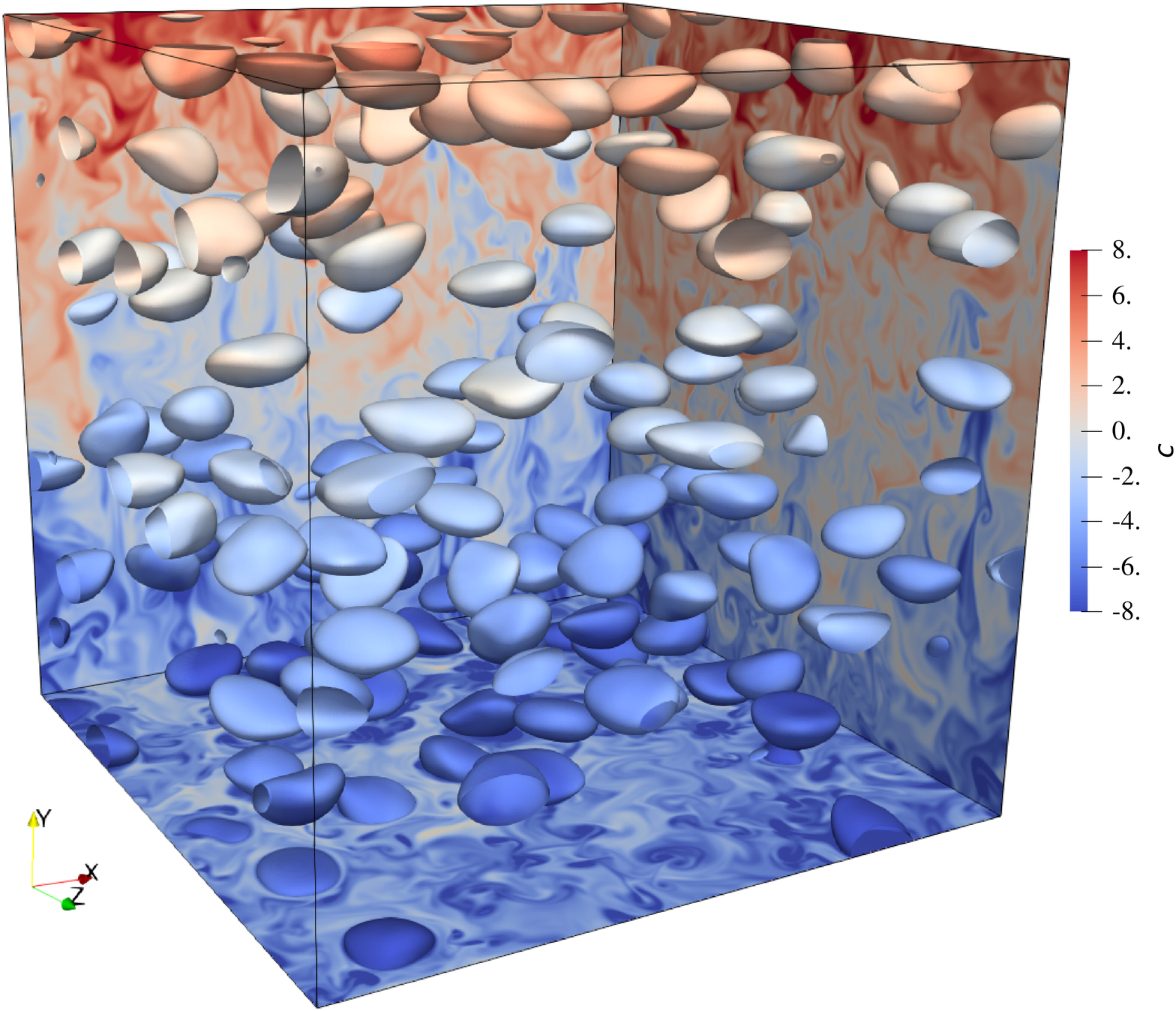}
    \end{overpic}
  \end{subfigure}
  \quad
  \begin{subfigure}[b]{0.48\linewidth}
    \caption{}
    \label{fig:c_tot_3D_hor}
    \begin{overpic}[trim = .3cm .3cm 0.3cm 0.3cm,clip,width=\linewidth]{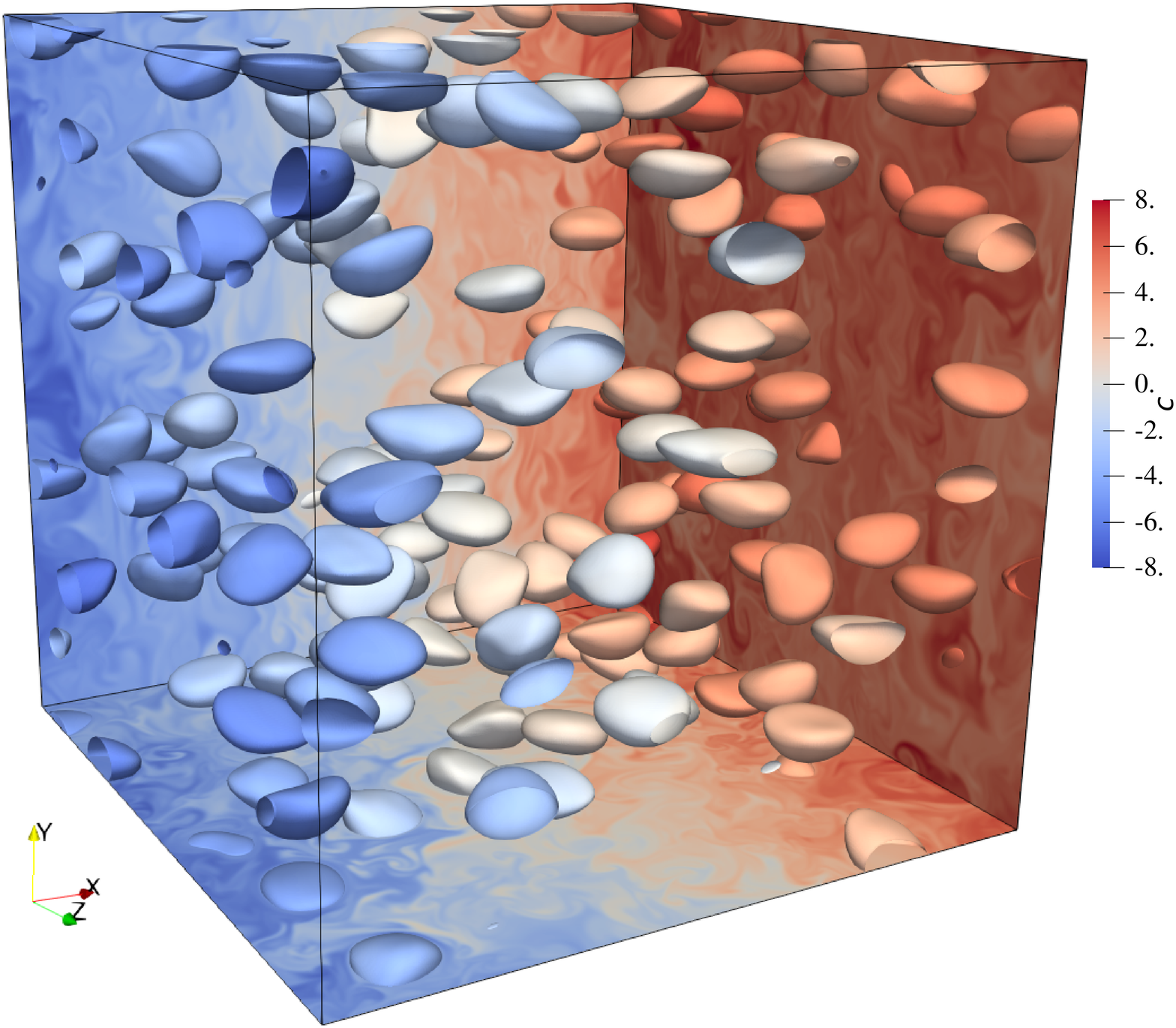}
    \end{overpic}
  \end{subfigure}
    \caption{Istantaneous snapshot from  case B4 with bubble interfaces and periodic domain boundaries coloured by the total scalar field $c$. The volume fraction is $\phi=5.2\%$, $Sc_l = 7$ and $\nabla^v \langle c \rangle=1$ is imposed in panel (a) and $\nabla^h \langle c \rangle=1$ on panel (b).}
    \label{fig:c_tot_3D}
\end{figure}

\begin{figure}[h]
    \centering
  \begin{subfigure}[b]{0.48\linewidth}
    \caption{}
    \label{fig:ca_tot_cross_sec}
    \begin{overpic}[trim = .3cm .3cm 0.3cm 0.3cm,clip,width=\linewidth]{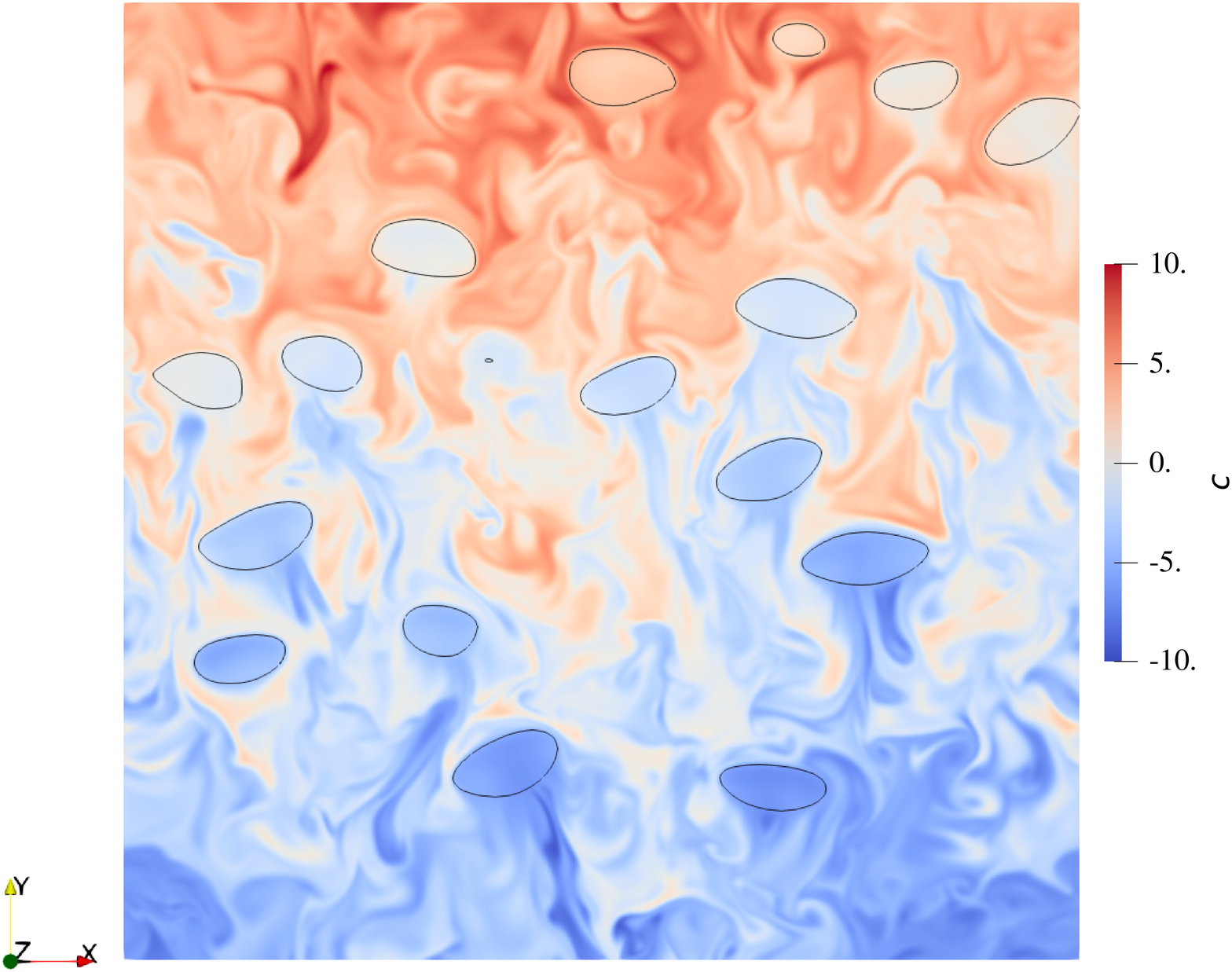}
    \end{overpic}
  \end{subfigure}
  \quad
  \begin{subfigure}[b]{0.48\linewidth}
    \caption{}
    \label{fig:ca_prim_cross_sec}
    \begin{overpic}[trim = .3cm .3cm 0.3cm 0.3cm,clip,width=\linewidth]{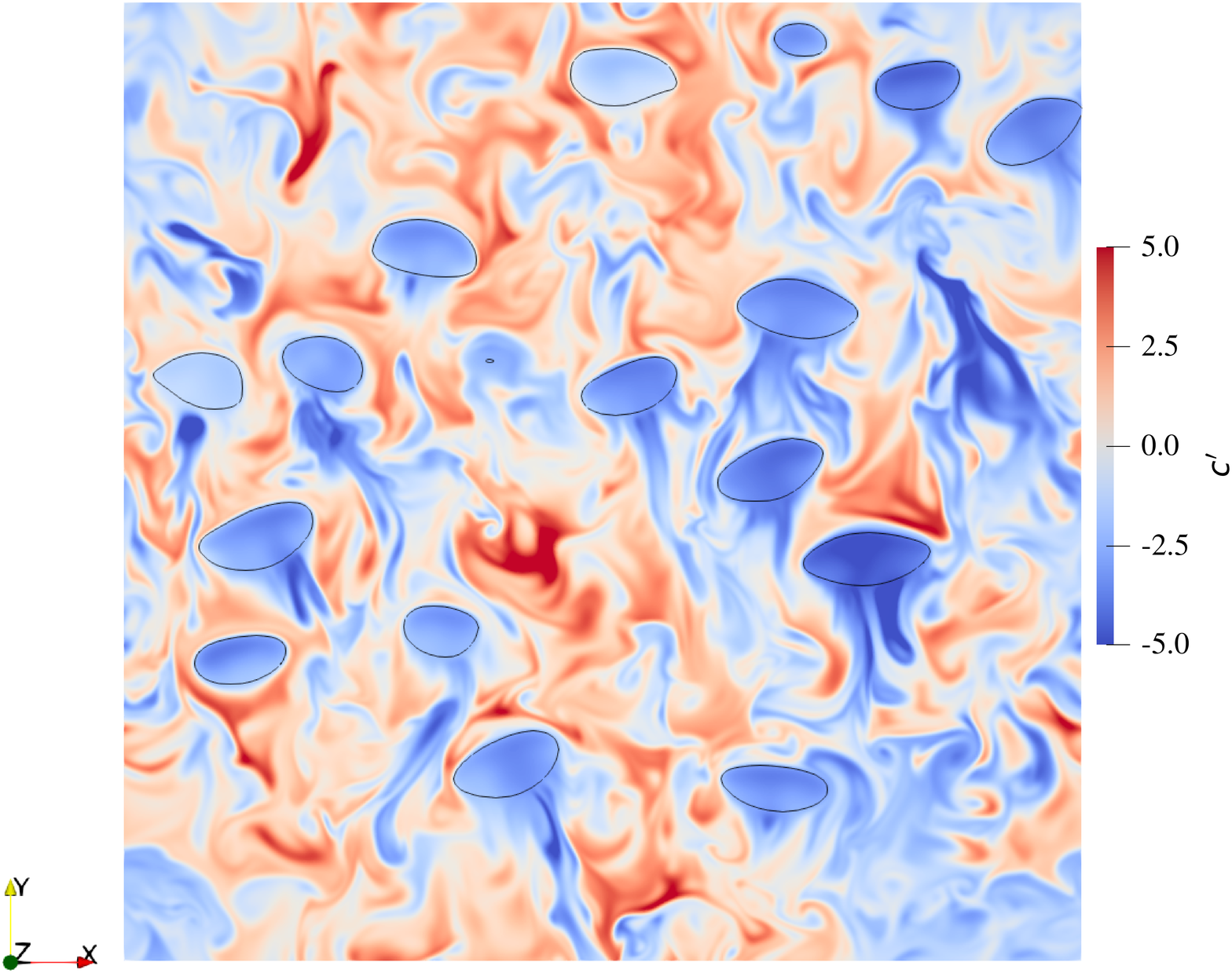}
    \end{overpic}
  \end{subfigure}
    \caption{Vertical cross-sectional view of the total $c$ and disturbance $c'$ scalar fields from case B4 with $\phi=5.2\%$, $\nabla^v \langle c \rangle=1$ and $Sc_l = 7$. The thin black lines represent the bubble interfaces.}
    \label{fig:ca_cross}
\end{figure}

\begin{figure}[h]
    \centering
  \begin{subfigure}[b]{0.48\linewidth}
    \caption{}
    \label{fig:cb_tot_cross_sec}
    \begin{overpic}[trim = .3cm .3cm 0.3cm 0.3cm,clip,width=\linewidth]{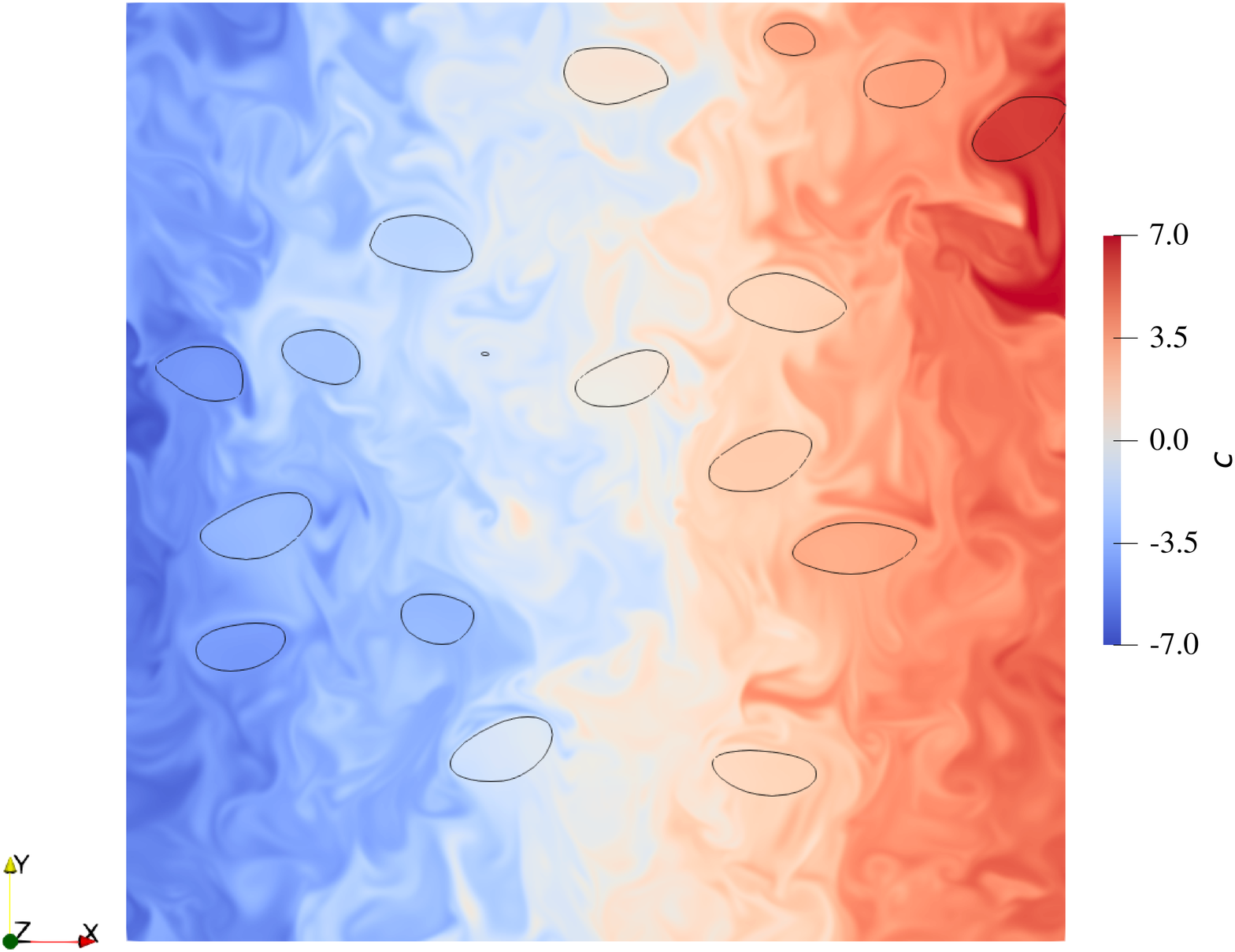}
    \end{overpic}
  \end{subfigure}
  \quad
  \begin{subfigure}[b]{0.48\linewidth}
    \caption{}
    \label{fig:cb_prim_cross_sec}
    \begin{overpic}[trim = .3cm .3cm 0.3cm 0.3cm,clip,width=\linewidth]{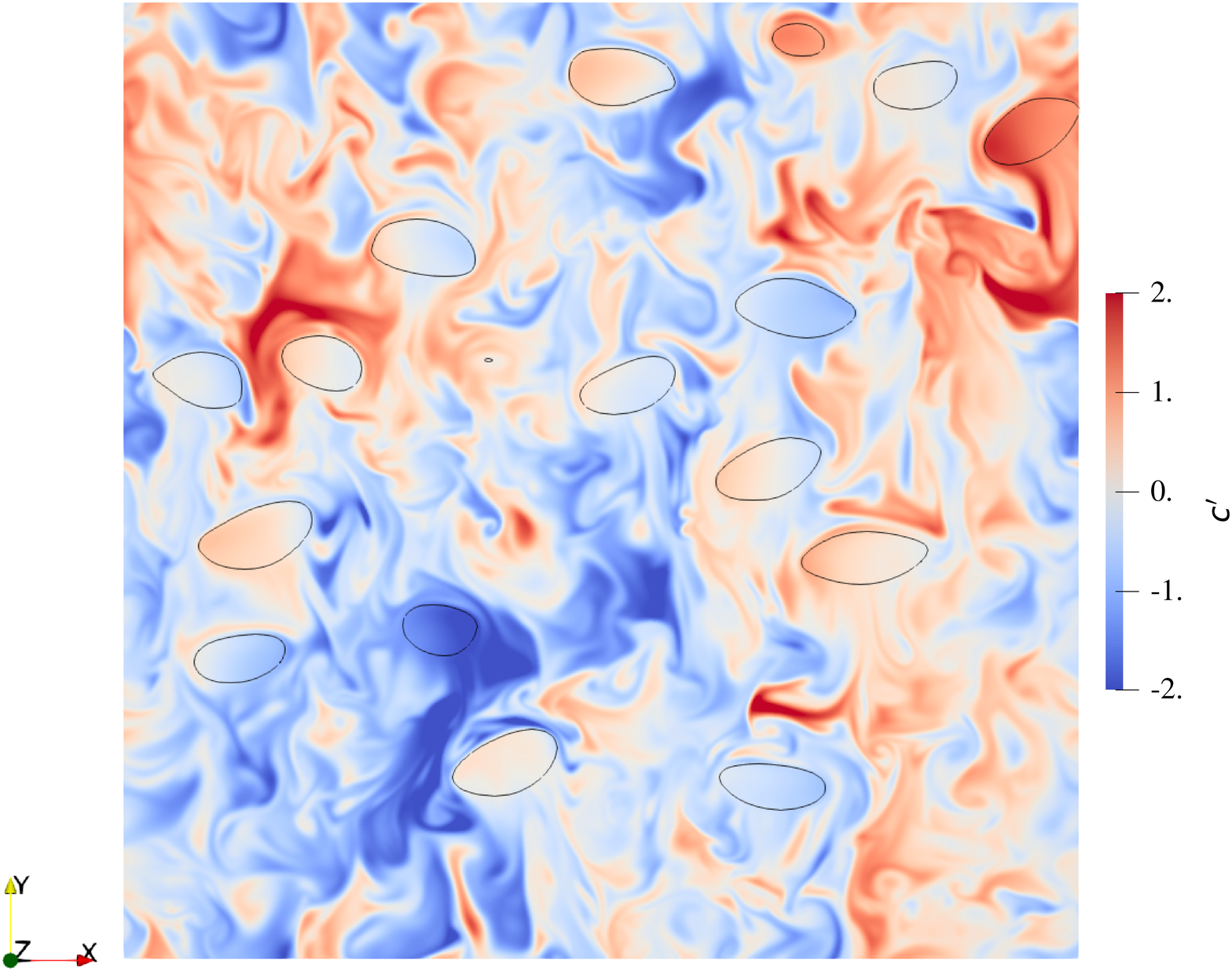}
    \end{overpic}
  \end{subfigure}
    \caption{Vertical cross-sectional view of the total $c$ and disturbance $c'$ scalar fields from case B4 with $\phi=5.2\%$, $\nabla^h \langle c \rangle=1$ and $Sc_l = 7$.}
    \label{fig:cb_cross}
\end{figure}

\Cref{fig:c_tot_3D} shows instantaneous snapshots from case B4 of the full periodic computational domain and the bubble interfaces coloured by the total scalar field $c$ with $\nabla^v \langle c \rangle$ in \cref{fig:c_tot_3D_vert} and $\nabla^h \langle c \rangle$ in \cref{fig:c_tot_3D_hor}. Vertical cross-sectional views of the total scalar field $c$ and the scalar disturbance field $c'$ from \cref{fig:c_tot_3D_vert} are shown in \cref{fig:ca_cross} and from \cref{fig:c_tot_3D_hor} in \cref{fig:cb_cross}.

The generation of the scalar disturbances stems from the last term on the r.h.s. of \cref{eq:scalar_transport} that (given a positive $\nabla^v \langle c \rangle$) predicts large negative disturbances in high upward-velocity regions and positive disturbances in downward velocity regions. Consequently, we speculate that the scalar disturbances in the case of $\nabla^v \langle c \rangle$ (shown in \cref{fig:ca_prim_cross_sec}) are generated by several interacting mechanisms. The first is the generation by the gas phase that preferentially moves in the vertical direction (generation at scales $k/k_{d_b}\approx1$). The second is the associated generation by the upward flowing liquid in the bubble wakes that is related to the capture-release mechanism described in \cite{Almeras2015, Almeras2018} (generation at scales $k/k_{d_b} \gtrapprox 1$). The third is the generation by the average downward flowing liquid in between the rising bubbles due to continuity (generation at scales $k/k_{d_b} \lessapprox 1$). The last is the production by the bubble-induced turbulence that was found to dominate the dispersion of a low-diffusive dye in a homogeneous bubbly flow similar to the flow considered in the present study \citep{Almeras2015}. 

In contrast, the scalar disturbances in the case of $\nabla^h \langle c \rangle$ (shown in \cref{fig:cb_prim_cross_sec}) show different characteristics. Here, the scalar disturbances are mainly generated by the bubble-induced turbulence since the average bubble motion in the horizontal directions equals zero. However, since the bubble-induced turbulence is characterised by anisotropy (due to the preferential motion of the bubble in the vertical direction), the influence of the bubble-induced turbulence on the scalar dynamics is not, in general, the same in cases with different mean gradient orientations. Given the different characteristics of the scalar disturbances in the cases of $\nabla^v \langle c \rangle$ and $\nabla^h \langle c \rangle$, we thus expect some differences in the scalar statistics that are analysed in \cref{sec:scalar_spectra}.

\subsection{Bubble clustering}

Following the methodology of \cite{tagawa2013} and \cite{Pandey2020}, we analyse the bubble clustering in our DNS with the volume fractions $\phi=1.7\%$ and $\phi=5.2\%$. We compute the centre-of-mass of each bubble in the DNS at a statistically steady state and compute Voronoi tesselations using the Voro++ library \citep{Rycroft2009}. For each volume fraction, we compute the standard deviation $\Sigma$ of the Voronoi volumes. This $\Sigma$ is compared to the standard deviation $\Sigma_{\mathit{rnd}}$ of the Voronoi volumes in 200 configurations of randomly positioned, non-overlapping bubbles in the same cubic domain with length $L$ as used in the DNS. \cite{tagawa2013} shows that the clustering indicator $\mathcal{C}=\Sigma/\Sigma_{\mathit{rnd}}$ quantitatively identifies different clustering morphologies. A value of $\mathcal{C}<1$ indicates a regular lattice
arrangement, $\mathcal{C}=1$ a random bubble distribution, and $\mathcal{C}>1$ irregular clustering. For our case with $\phi=1.7\%$, we obtain $\mathcal{C}=0.95$ and for $\phi=5.2\%$ we have $\mathcal{C}=1.2$. These values indicate random or weakly irregular clustering in our DNS \citep{Pandey2020}. Consequently, we observe no significant effect of clustering on our results.

\subsection{Velocity spectra}

We define the velocity spectrum $E_u$ of the suspension at wavenumber $k=\lvert{\bf k}\rvert$ as
\begin{equation}
     E_u(k) =\frac{1}{2} \sum_{k-\Delta k/2 < \lvert{\bf k}\rvert < k+\Delta k/2} \dfrac{\langle \hat{\boldsymbol{u}}'({\bf k}) {\boldsymbol{\cdot}} \hat{\boldsymbol{u}}'^*{(\bf k}) \rangle}{\Delta k}\ , \label{eq:Eu}
\end{equation}
where \ ' $\hat{ }$\ ' represents the Fourier mode, '$^*$' is the complex conjugate and $\langle \cdot \rangle$ is the ensemble average. 
Similarly, we can define the velocity spectrum of the liquid phase considering the liquid velocity ${\boldsymbol{\tilde{u}_l}=\boldsymbol{u}\lvert{(f=1)}}$; however, we regularise the liquid velocity directly with the volume fraction field to avoid discontinuities that can lead to possible Gibbs phenomena in the Fourier transform defining the regularised liquid velocity as $\boldsymbol{u_l}=f\boldsymbol{u}$:
\begin{equation}
     {E}_{u,l}(k) =\frac{1}{2} \sum_{k-\Delta k/2 < \lvert{\bf k}\rvert < k+\Delta k/2}\dfrac{ \langle \hat{\boldsymbol{u_l}}'({\bf k}) {\boldsymbol{\cdot}} \hat{\boldsymbol{u_l}}'^*{(\bf k}) \rangle}{\Delta k}\ .\label{eq:Eu_l}
\end{equation}

\Cref{fig:spectra_vel} shows the suspension and liquid velocity spectra for our cases, and previous bubbly flow DNS by \cite{Pandey2020} at similar governing parameters (case R7 in that study). All cases show a peak at scales close to the bubble diameter ($k/k_{d_b}\approx1$, with $k_{d_b}=2\pi/d_b$) whereas at $k/k_{d_b} > 1$, the velocity spectra of the bubbly flows scale approximately as $k^{-3}$ in agreement with previous studies. The single-phase velocity spectra show an even steeper slope since, here, the velocity fluctuations are only produced at the forcing length scale. his is opposed to bubbly flows where velocity fluctuations are produced by both large scales, and at $k/k_{d_b} > 1$ where the fluctuations are continuously produced and directly dissipated in the bubble wakes \citep{lance_bataille_1991}. The single-phase velocity spectrum does not show the Kolmogorov -5/3 scaling in a large interval since the Taylor-Reynolds number is only 39.
The $-3$ scaling in the bubbly flows is more pronounced in the liquid than in the suspension, so we focus on the scalar dynamics in the liquid phase.

\subsection{Scalar spectra} \label{sec:scalar_spectra}

We define the suspension and liquid scalar spectra similar to \cref{eq:Eu} and \cref{eq:Eu_l}:
\begin{equation}
     E_c(k) = \sum_{k-\Delta k/2 < \lvert{\bf k}\rvert < k+\Delta k/2} \dfrac{\langle \hat{{c}}'({\bf k})  \hat{{c}}'^*{(\bf k}) \rangle}{\Delta k}, \; \; E_{c,l}(k) = \sum_{k-\Delta k/2 < \lvert{\bf k}\rvert < k+\Delta k/2} \dfrac{\langle \hat{{c_l}}'({\bf k})  \hat{{c_l}}'^*{(\bf k}) \rangle}{\Delta k}\ .
\end{equation}

\begin{figure}[h]
\centering
  \begin{subfigure}[b]{0.48\linewidth}
   \caption{}
    \label{fig:spectra_vel}
    \begin{overpic}[trim = .3cm .3cm 0.3cm 0.3cm,clip,width=\linewidth]{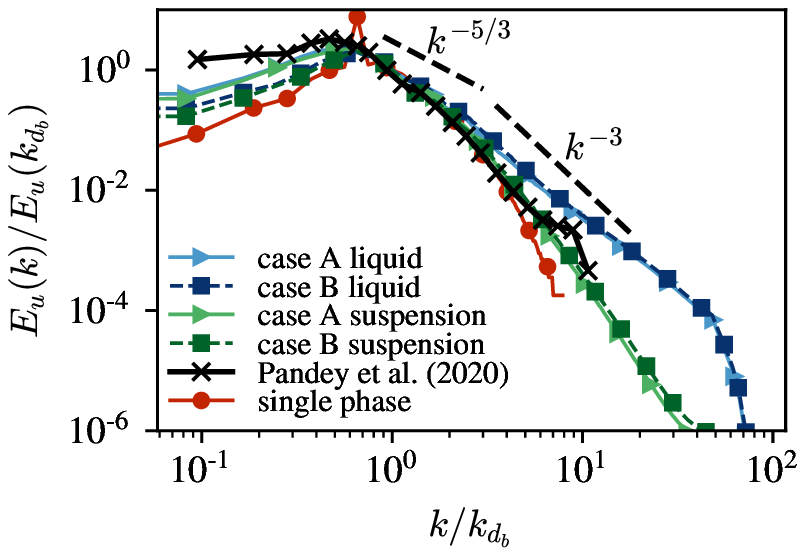}
    \end{overpic}
  \end{subfigure}
  \quad
  \begin{subfigure}[b]{0.48\linewidth}
    \caption{}
    \label{fig:spectra_scalar_single}
    \begin{overpic}[trim = .3cm .3cm 0.3cm 0.3cm,clip,width=\linewidth]{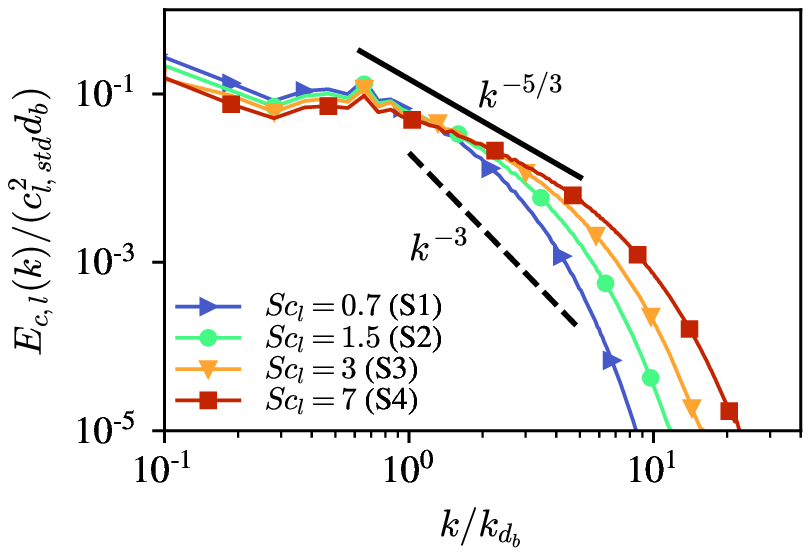}
    \end{overpic}
  \end{subfigure}
    \begin{subfigure}[b]{0.48\linewidth}
  \centering
    \caption{}
    \label{fig:spectra_vert}
    \begin{overpic}[trim = .3cm .3cm 0.3cm 0.3cm,clip,width=\linewidth]{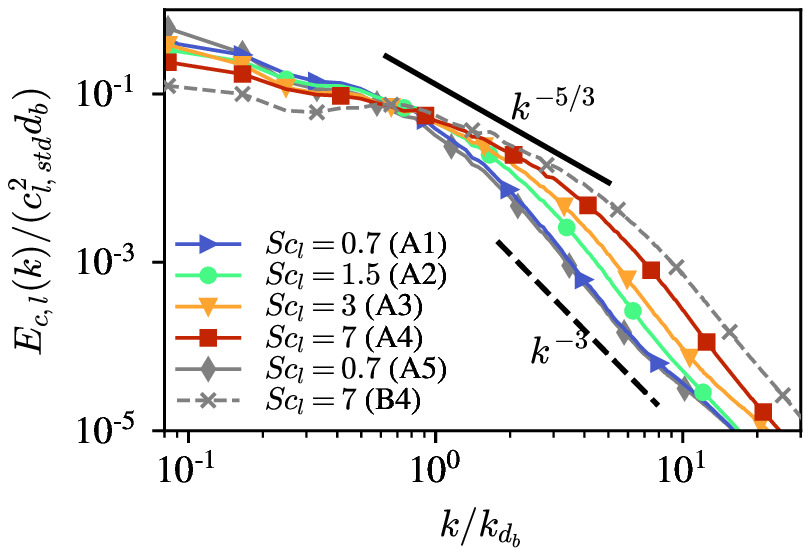}
    \end{overpic}
  \end{subfigure}
  \quad
  \begin{subfigure}[b]{0.48\linewidth}
  \centering
    \caption{}
    \label{fig:spectra_hor}
    \begin{overpic}[trim = .3cm .3cm 0.3cm 0.3cm,clip,width=\linewidth]{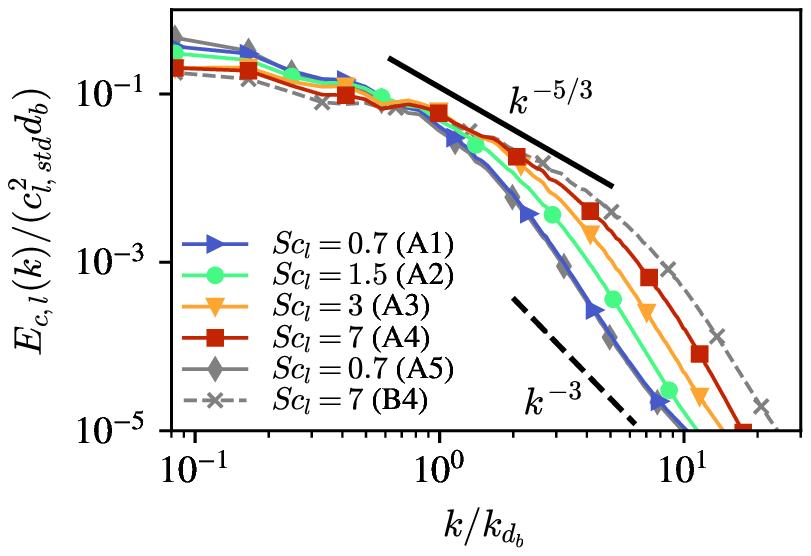}
    \end{overpic}
  \end{subfigure}
    \label{fig:vel_and_single_spectra}
    \caption{Normalised velocity spectra $E_u(k)$ and scalar spectra $E_{c,l}(k)$  against the normalised wave number $k/k_{d_b}$. (a): Velocity spectra of single-phase, suspensions and liquid phases at two different volume fractions and comparisons with \cite{Pandey2020}. (b): $E_c(k)$ for the single phase isotropic turbulence case. (c): Bubbly flow cases with $\nabla^v \langle c \rangle$. (d): Bubbly flow cases with $\nabla^h \langle c \rangle$. The gas volume fraction is $\phi=1.7\%$ in all cases A1 - A5. Case B4 with $\phi=5.2\%$ is included to illustrate the effects of $\phi$ on $E_{c,l}(k)$.}
\end{figure}

The single-phase scalar spectra are shown in \cref{fig:spectra_scalar_single}. Below the forcing scale, the classical $-5/3$ scaling emerges, especially for the highest $Sc$-numbers. For lower $Sc$-numbers, the diffusion significantly influences the dynamics below the forcing scale, like in the velocity spectra, decreasing exponentially due to viscous dissipation \citep{pope2001}.

The normalised liquid scalar spectra for the bubbly flow simulations are shown in \cref{fig:spectra_vert} for the cases with $\nabla^v \langle c \rangle$ and in \cref{fig:spectra_hor} for the cases with $\nabla^h \langle c \rangle$. For brevity, we only show the spectra for all cases at $\phi=1.7\%$ since the general trends are the same at $\phi=5.2\%$. Case B4 is, however, included to illustrate the minor effects of a higher $\phi=5.2\%$ on $E_c/(c^2_{\mathit{std}}d_b)$. We do not see the different scaling laws at different volume fractions like in the experiments by \cite{dung2022emergence}. However, in that experiments, the liquid turbulence results from the interaction between the bubble-induced turbulence and the incident turbulence generated by an active grid. This experimental system corresponds to a finite value of the bubblance parameter (ratio of the kinetic energy produced by the bubbles to the turbulent kinetic energy in the absence of bubbles, \cite{rensen2005}). This is in contrast with the present DNS results with only bubble-induced turbulence and an infinite bubblance parameter. Indeed, \cite{prakash2016} showed experimentally how the $-3$ scaling of the velocity spectra emerged as the bubblance parameter was modified. This difference between the present DNS and the experiment by \cite{dung2022emergence} is a possible reason why we do not observe different scaling laws at different volume fractions. In addition, the latter experimental study considers a thermal mixing layer with a nonlinear mean gradient, whereas the present study imposes a constant mean scalar gradient. The influence of a nonlinear gradient on the scalar spectrum is not clear but may also explain the different observations.

In \cref{fig:spectra_vert} and \cref{fig:spectra_hor}, we observe a maximum of the scalar spectra at the lowest wavenumber, a signature that the scalar is directly forced by the mean gradients. Consistent with the results of \cite{dung2022emergence}, the liquid scalar spectra show two different scalings. 
The larger scales show a standard $-5/3$ scaling only in a narrow range of wavenumbers. However, after a transition scale, the spectra scaling changes towards the -3 scaling or lower for all the Schmidt numbers.
When the gradient is in the horizontal direction (\cref{fig:spectra_hor}), we observe a faster decay of the scalar spectrum below the bubble diameter wavenumber. The transition wavenumber increases with the Schmidt number, depending on the molecular diffusivity. It coincides approximately with the bubble wavenumber for the smaller $Sc$-numbers since, at $Sc$-numbers of order one, we expect that the dynamics of the passive scalar transport are similar to the momentum transport. Case A5 is included to assess the effects of increasing $Sc_g$ on the scalar dynamics while keeping the $Sc_l=0.7$ as in case A1; this case corresponds to the same scalar diffusivity in the gas and liquid phases. We observe that also, in this case, the normalised spectrum shows a -3 scaling after the bubble diameter wavenumber. 
\Cref{fig:spectra_vert} and \cref{fig:spectra_hor} show only minor differences between cases A1 and A5, indicating that the $Sc_g$-number does not significantly influence the liquid scalar fluctuations. This is, however, not true for the statistics of the suspension, as discussed later.

\begin{figure}[h]
\centering
  \begin{subfigure}[b]{0.48\linewidth}
  \caption{}
    \label{fig:comp_single}
    \begin{overpic}[trim = .3cm .3cm 0.3cm 0.3cm,clip,width=\linewidth]{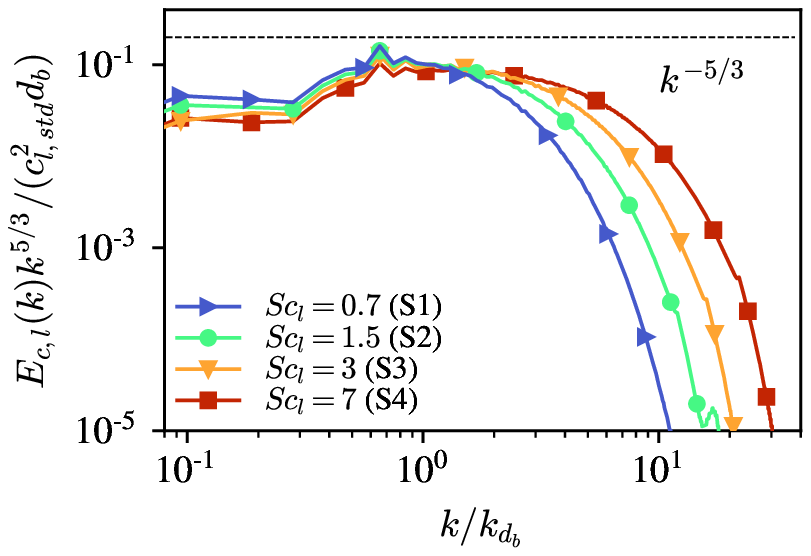}
    \end{overpic}
  \end{subfigure}
  \quad
  \begin{subfigure}[b]{0.48\linewidth}
    \caption{}
    \label{fig:comp_hor53}
    \begin{overpic}[trim = .3cm .3cm 0.3cm 0.3cm,clip,width=\linewidth]{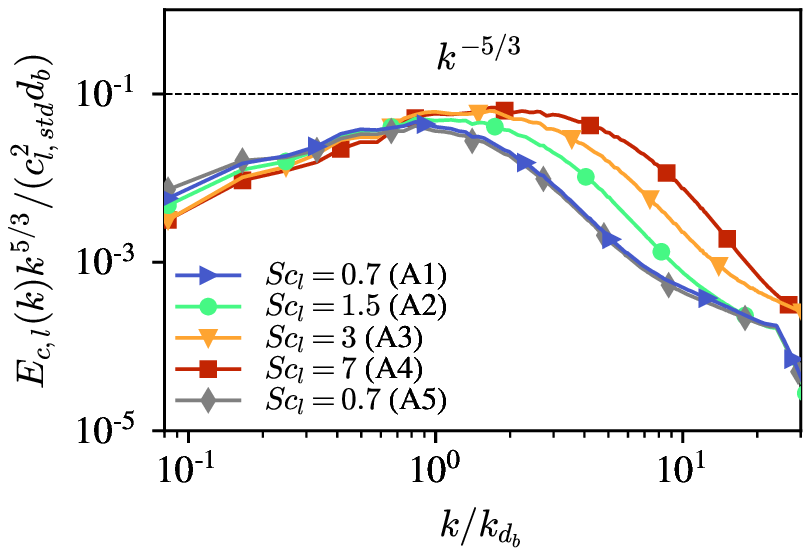}
    \end{overpic}
  \end{subfigure}
    \begin{subfigure}[b]{0.48\linewidth}
  \centering
  \caption{}
    \label{fig:comp_vert3}
    \begin{overpic}[trim = .3cm .3cm 0.3cm 0.3cm,clip,width=\linewidth]{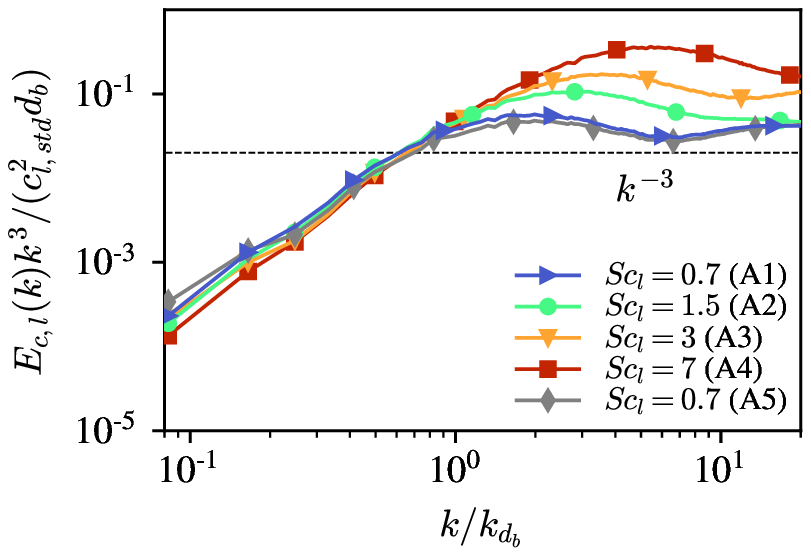}
    \end{overpic}
  \end{subfigure}
  \quad
    
    \caption{Compensated and normalised scalar spectra $E_{c,l}(k)$ against the normalised wave number $k/k_{d_b}$. (a): Single phase isotropic turbulence cases compensated by $k^{5/3}$. (b): Bubbly flow cases with $\nabla^h \langle c \rangle$ compensated by $k^{5/3}$ (c): Bubbly flow cases with $\nabla^v \langle c \rangle$ compensated by $k^{3}$.}
    \label{fig:comp_scalar_spectra}
\end{figure}

\Cref{fig:comp_scalar_spectra} shows the compensated scalar spectra to more clearly illustrate the scalar spectra scalings and transitions. \Cref{fig:comp_single} shows that the classical $-5/3$ scaling in the single phase cases emerges in a narrow range of wavenumbers (where the range increase with the $\mathit{Sc}_l$-number) after the forcing length scale. This limited range is due to an insufficient Taylor-P{é}clet number. Defining the Taylor length scale for the scalar field as $\lambda_c = \sqrt{6D_l/\varepsilon_c}c_{l,\mathit{std}}$ we obtain Taylor-P{é}clet numbers $\mathit{Pe}_\lambda = u_{l,\mathit{std}} \lambda_c / D_l$ in the range of 33-143 and the Taylor length scales $\lambda_c$ in the range of 0.15-0.6$d_b$. These values indicate that molecular diffusion significantly affects the scalar dynamics at scales comparable to or slightly smaller than the bubble size $d_b$. It is thus reasonable that the inertial range with the $-5/3$ scaling is not very pronounced.
The $-5/3$ scaling is, however, well established for single-phase DNS with wider inertial ranges \citep{Corrsin1951, Gotoh2012}. Since the forcing length scale in the single phase cases is close to the bubble size, it is reasonable that the $-5/3$-scaling is not pronounced in the bubbly flow cases either, as shown in the compensated spectra of \cref{fig:comp_hor53}.

\Cref{fig:comp_vert3} shows the compensated scalar spectra for the bubble flow cases with $\nabla^v \langle c \rangle$. Here it is clear that the spectra change scaling towards approximately $-3$ at scales $k/k_{d_b} \geq 1$ and that the transition occurs at higher $k$ for increasing $\mathit{Sc}_l$.

\subsection{Scalar spectral budget}

The transition to the -3 scaling is clearly a footprint of the bubble-induced agitation mechanisms that influence the scalar dynamics also at scales below the bubble diameter $k/k_{d_b}\gg1$. To investigate the origin of this scaling, we compute the spectral budget of the passive scalar by manipulating \cref{eq:scalar_transport} in Fourier space \citep{monin1975}:
\begin{equation}
    \dfrac{\partial E_c(k,t)}{\partial t} = T_c(k,t) - \mathcal{E}_c(k,t) + P_c(k,t)\, , \label{eq:scalar_budget}
\end{equation}
where  $T_c(k,t) = \sum_{(k-\Delta k/2 < \lvert{\bf k}\rvert < k+\Delta k/2)} -\langle \hat{h_c}\hat{c}'^* + \hat{h_c}^*\hat{c}' \rangle / \Delta k$ is the local transfer term, $\mathcal{E}_c(k,t) = \sum_{(k-\Delta k/2 < \lvert{\bf k}\rvert < k+\Delta k/2)} \langle \hat{d_c}\hat{c}'^* + \hat{d_c}^*\hat{c}' \rangle / \Delta k$ is the dissipation term and $P_c(k,t) = \sum_{(k-\Delta k/2 < \lvert{\bf k}\rvert < k+\Delta k/2)}  \nabla \langle c \rangle \boldsymbol{\cdot} \langle(\hat{\boldsymbol{u}}'\hat{c}'^* + \hat{\boldsymbol{u}}'^*\hat{c}')\rangle / \Delta k$ is the production at wavenumber $k$.

\cite{dung2022emergence} speculated that the passive scalar transport mechanisms are not due to direct scalar production and simultaneous dissipation at scales smaller than bubble diameter, as for the momentum transport. Instead, scalar production, generated at larger scales by the mean gradient, decays faster so that at small scales just a balance between liquid scalar transfer $T_{c,l}$ and spectral dissipation $\mathcal{E}_{c,l}(k,t)$ occurs. By dimensional arguments, they show that if the liquid scalar transfer is only a function of scalar dissipation and wavenumber, then the scaling is a -1 power-law for the wavenumber: $T_{c,l}=T_{c,l}(\varepsilon_{c,l},k)\propto \varepsilon_{c,l}k^{-1}$.
Since the liquid spectral dissipation is defined as $\mathcal{E}_{c,l}(k,t)=2D_lk^2E_{c,l}$, it is trivial to show that if a balance exists with the scalar transfer term, then the scalar spectra should scale as a -3 power-law in $k$, $E_{c,l} \propto k^{-3}$.
The different terms of \cref{eq:scalar_budget} are challenging to measure in experiments, so, as suggested in \cite{dung2022emergence}, DNS simulations are needed to estimate them. Here, we assess their hypotheses, extracting the relevant statistics from our numerical simulations.

\begin{figure}
  \centering
  \begin{subfigure}[b]{0.335\linewidth}
  \centering
 \caption{}
    \label{fig:ProdSingle}
    \begin{overpic}[trim = .3cm .3cm 0.3cm 0.3cm,clip,width=\linewidth]{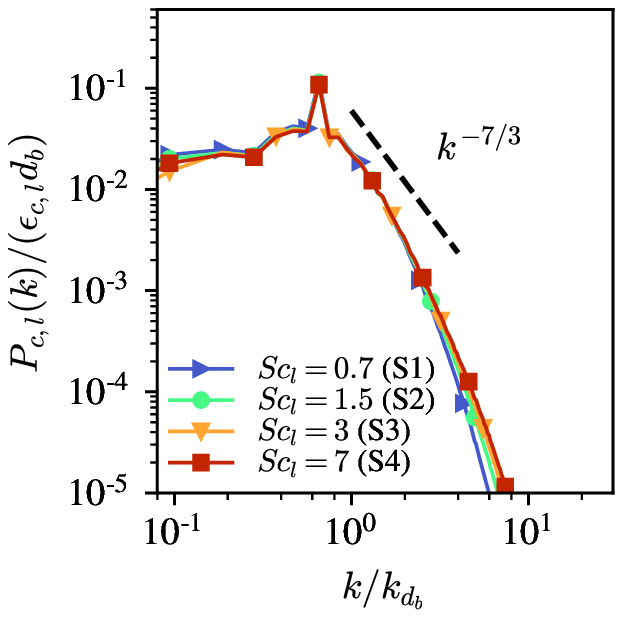}
    \end{overpic}
  \end{subfigure}
  \quad
  \begin{subfigure}[b]{0.3\linewidth}
  \centering
   \caption{}
    \label{fig:ProdAvert}
    \begin{overpic}[trim = .9cm .3cm 0.3cm 0.3cm,clip,width=\linewidth]{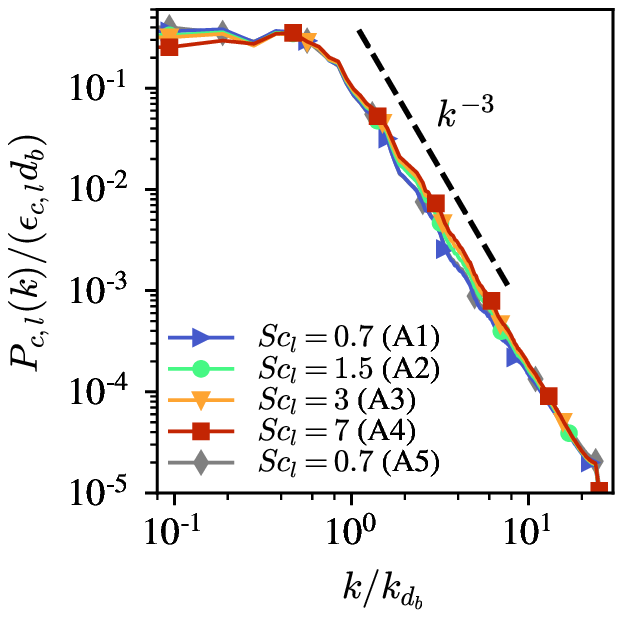}
    \end{overpic}
  \end{subfigure}
    \quad
  \begin{subfigure}[b]{0.3\linewidth}
  \centering
\caption{}
    \label{fig:ProdAhori}
    \begin{overpic}[trim = .9cm .3cm 0.3cm 0.3cm,clip,width=\linewidth]{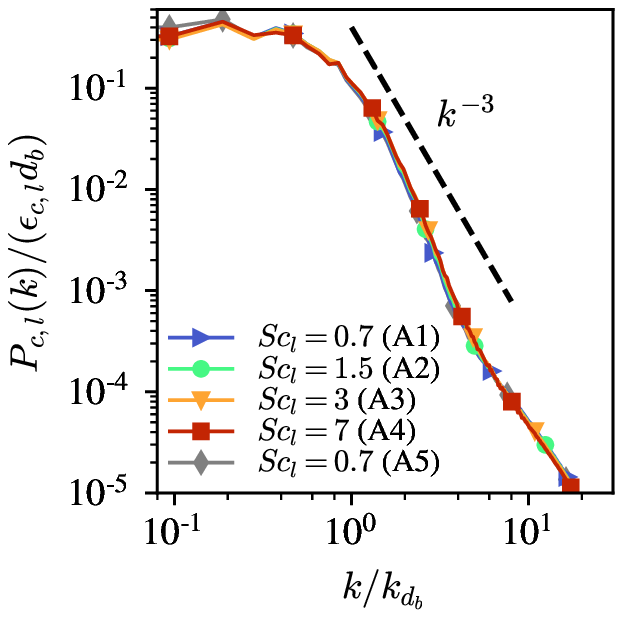}
    \end{overpic}
  \end{subfigure}
  
  \begin{subfigure}[b]{0.335\linewidth}
  \centering
  \caption{}
    \label{fig:TransSingle}
    \begin{overpic}[trim = .3cm .3cm 0.3cm 0.3cm,clip,width=\linewidth]{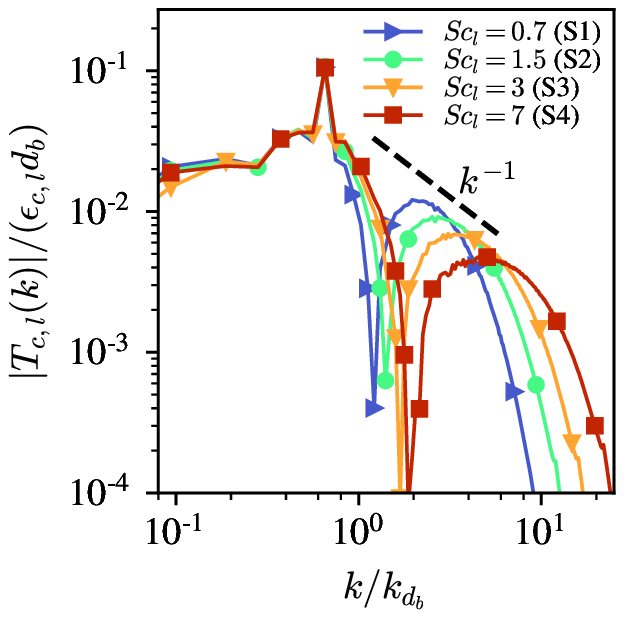}
    \end{overpic}
  \end{subfigure}
  \quad
  \begin{subfigure}[b]{0.3\linewidth}
  \centering
\caption{}
    \label{fig:TransAvert}
    \begin{overpic}[trim = 0.9cm .3cm 0.3cm 0.3cm,clip,width=\linewidth]{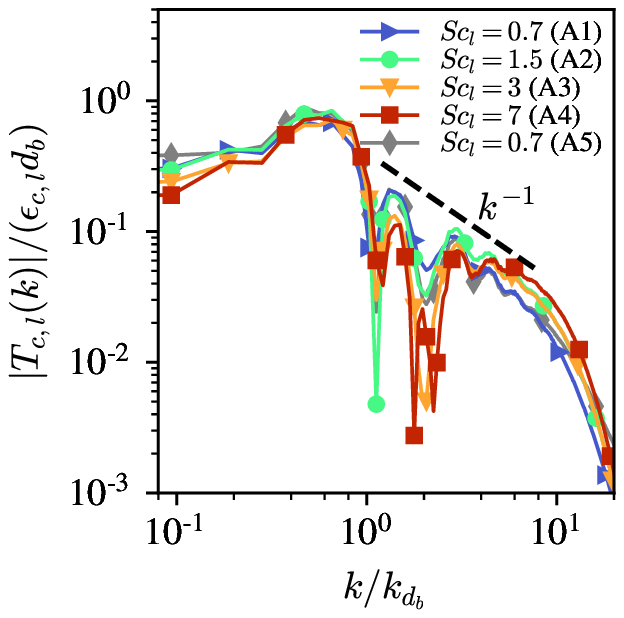}
    \end{overpic}
  \end{subfigure}
    \quad
  \begin{subfigure}[b]{0.3\linewidth}
  \centering
  \caption{}
    \label{fig:TransAhori}
    \begin{overpic}[trim = .9cm .3cm 0.3cm 0.3cm,clip,width=\linewidth]{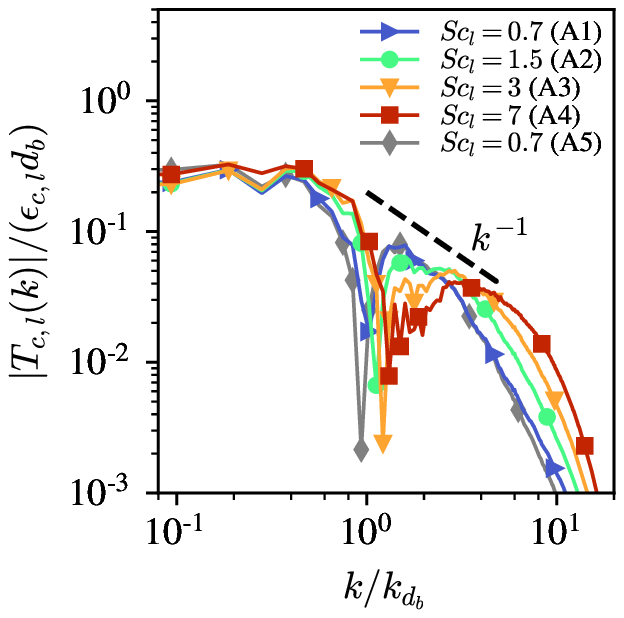}
    \end{overpic}
  \end{subfigure}
  
    \caption{Normalised scalar production term $P_{c,l}(k)/(\epsilon_{c,l} d_b)$ for the single phase cases (a) and bubbly flow cases A1-A5 with $\phi=1.7\%$ for $\nabla^v \langle c \rangle$ (b) and $\nabla^h \langle c \rangle$ (c). Normalised scalar transfer term $|T_{c,l}(k)|/(\epsilon_{c,l} d_b)$ for the corresponding single-phase cases (d) and bubbly flow cases with $\nabla^v \langle c \rangle$ (e) and $\nabla^h \langle c \rangle$ (f).}
    \label{fig:prod}
\end{figure}

In the top panels of \cref{fig:prod}, we show the normalised production $P_{c,l}(k)/(\epsilon_{c,l} d_b)$ of \cref{eq:scalar_budget} for the isotropic single phase $(a)$ and bubbly flow cases with $\nabla^v \langle c \rangle$ $(b)$ and $\nabla^h \langle c \rangle$ $(c)$. The panels show that a significant portion of the scalar fluctuations is produced at about $k/k_{d_b} \leq 1$ for both the single phase and bubbly flow cases. At $k/k_{d_b}  > 1$, $P_{c,l}(k)$ scales as about $-7/3$ in a narrow range of $k$ for the single phase cases (as predicted by \cite{Lumley1964} with dimensional analysis and found in the DNS of \cite{Gotoh2012}). Bubble suspensions, instead, show that $P_{c,l}(k)$ decays with a faster power-law (about $k^{-3}$). Therefore, at a statistically steady state, \cref{eq:scalar_budget} implies that the scalar spectra $E_{c,l}(k)$ at large wavenumbers are governed by the balance of the net transfer and the diffusive dissipation. Interestingly, we observe a self-similarity in the normalised production spectra for all the cases at different liquid diffusivity. This self-similarity can probably be explained by the fact that the production at large scales is due to the mean scalar gradient, while the liquid diffusivity only influences the smaller scales where the production is small.

\begin{figure}
    \centering
    \includegraphics[width=12cm]{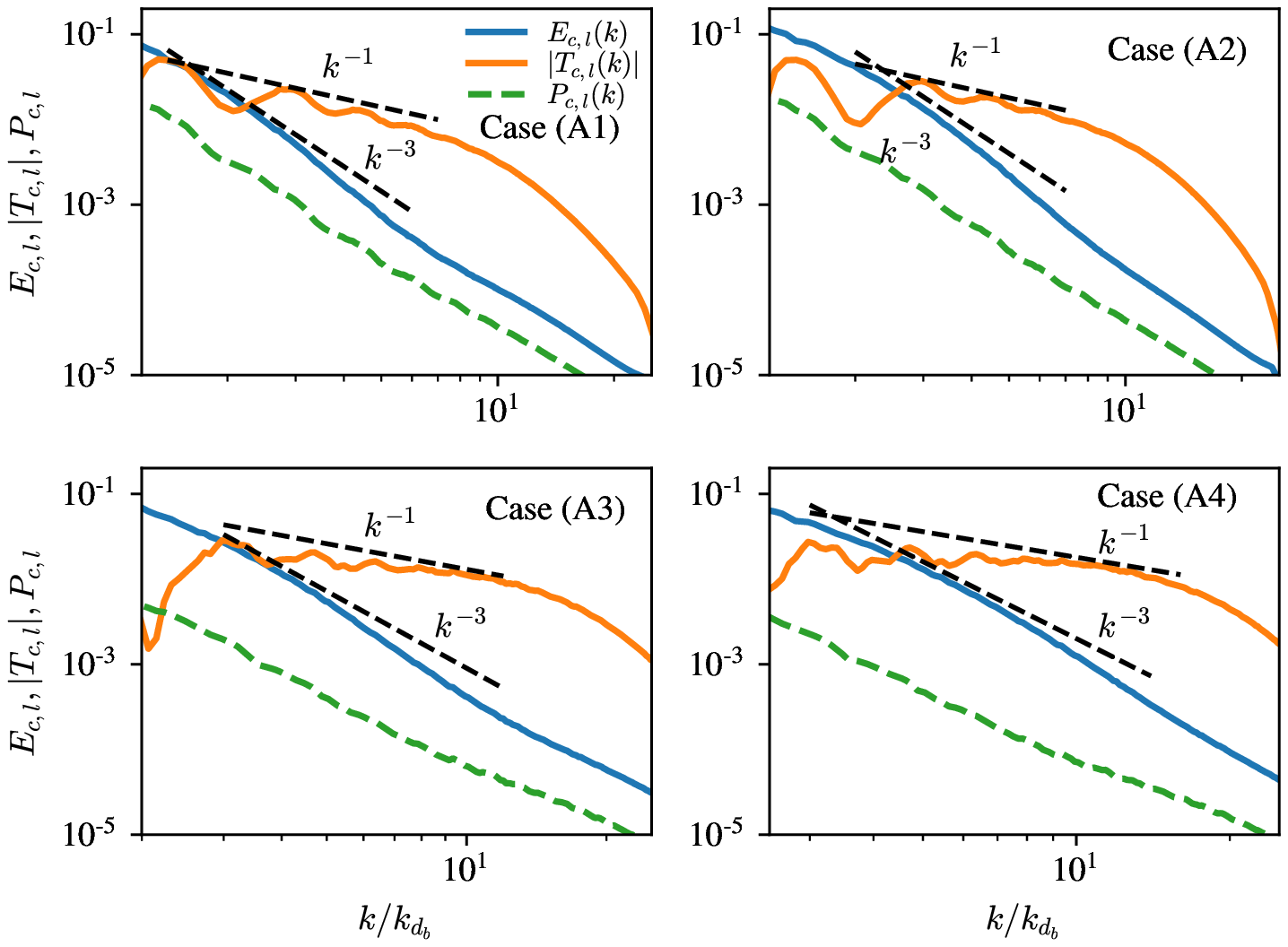}
    \caption{Zoomed-in plots of the scalar, transfer and production spectra for our cases A1-A4 with $\nabla^v \langle c \rangle$. In the same range where $|T_{c,l}(k)| \propto k^{-1}$, the scalar spectra shows a transition to approximately $E_{c,l} \propto k^{-3}$. 
    }
    \label{fig:budget_Av}
\end{figure}

The bottom panels of \cref{fig:prod} show the absolute value of the scalar transfer spectra. When the production becomes significantly smaller than the transfer, the latter scales close to $-1$ in a limited range of wavenumbers. In the same range, the scalar spectra show approximately a $-3$ scaling. These trends are highlighted in the zoomed-in plots of \cref{fig:budget_Av} that show the scalar spectrum, absolute transfer and the production spectrum for each case A1-A4 with $\nabla^v \langle c \rangle$. These plots are consistent with the proposed hypothesis in \citep{dung2022emergence} that, at large wavenumbers ($k/k_{d_b} \gtrapprox 3$ in \cref{fig:budget_Av}), the scalar spectrum is governed by the balance of the net transfer and the diffusive dissipation (as the production is negligible) and that the $-1$ scaling of the transfer term induces the $-3$ scaling of the scalar spectra. 
The trends shown in \cref{fig:budget_Av} are observed for all our bubble suspension cases.

\subsection{Effective scalar diffusivity} \label{sec:eff_diff}

The effective scalar diffusivity describes the macroscale scalar flux in the suspension seen as a continuum. Following the detailed derivation of \cite{Loisy_thesis}, the contributions to the effective diffusivity are obtained by ensemble-averaging the scalar transport equation:
\begin{equation}
    \dfrac{\partial \langle c \rangle }{\partial t} + \nabla \boldsymbol{\cdot} \langle \boldsymbol{q} \rangle = 0 \, , \label{eq:phase-av-c}
\end{equation}
where the average flux can be formulated as
\begin{equation}
    \langle \boldsymbol{q} \rangle = \langle\boldsymbol{u}\rangle \langle c\rangle - D_{\mathit{mol,s}} \nabla\langle c\rangle + (1-\phi)\langle \boldsymbol{u}'c' \rangle_l + \phi\langle \boldsymbol{u}'c' \rangle_g - (1-\phi)D_l \langle \nabla c' \rangle_l -\phi D_g \langle \nabla c' \rangle_g \, .
\end{equation}
Here, we have used that the phase-ensemble average of an arbitrary variable $A_s$ of the suspension is $\langle A_s \rangle = \langle f A_l + (1-f)A_g \rangle = (1-\phi)\langle A_l \rangle + \phi\langle A_g \rangle$. 
The effective diffusivity tensor is defined as
\begin{align}
    \bm{\mathsf{D}}_{\mathit{eff}} &= D_{\mathit{mol,s}} \bm{\mathsf{I}} + \bm{\mathsf{D}}_{\mathit{conv}} + \bm{\mathsf{D}}_{\mathit{diff}}\, ,\\
    \bm{\mathsf{D}}_{\mathit{conv}} \boldsymbol{\cdot} \nabla \langle c \rangle &= -(1-\phi)\langle \boldsymbol{u}'c' \rangle_l - \phi\langle \boldsymbol{u}'c' \rangle_g\, \label{eq:D_conv} ,\\
    \bm{\mathsf{D}}_{\mathit{diff}}\boldsymbol{\cdot} \nabla \langle c \rangle &= (1-\phi)D_l \langle \nabla c' \rangle_l +\phi D_g \langle \nabla c' \rangle_g\, ,
\end{align}
so that \cref{eq:phase-av-c} becomes
\begin{equation}
    \dfrac{\partial \langle c \rangle }{\partial t} + \nabla \boldsymbol{\cdot} (\langle\boldsymbol{u}\rangle \langle c\rangle) - \nabla \boldsymbol{\cdot} (\bm{\mathsf{D}}_{\mathit{eff}} \boldsymbol{\cdot} \nabla \langle c\rangle) = 0\, .
\end{equation}

For all cases in this study, $\bm{\mathsf{D}}_{\mathit{diff}} < 0.02\bm{\mathsf{D}}_{\mathit{conv}}$. The $\bm{\mathsf{D}}_{\mathit{conv}}$ is larger also in \cite{Loisy2018} for lower bubble Reynolds numbers ($Re\approx30$) and with the same diffusivity in both phases. We define $\mathsf{D}^v_{\mathit{conv}}$ as the non-zero diagonal component of $\bm{\mathsf{D}}_{\mathit{conv}}$ for an imposed scalar gradient $\nabla^v \langle c \rangle$ and $\mathsf{D}^h_{\mathit{conv}}$ for a $\nabla^h \langle c \rangle$. All off-diagonal components of $\bm{\mathsf{D}}_{\mathit{conv}}$ are zero \citep{Loisy_thesis}.

\begin{figure}
  \centering
  \begin{subfigure}[t]{0.33\linewidth}
  \centering
  \caption{}
    \label{fig:NuPe_std}
    \begin{overpic}[trim = .3cm .3cm 0.3cm 0.3cm,clip,width=\linewidth]{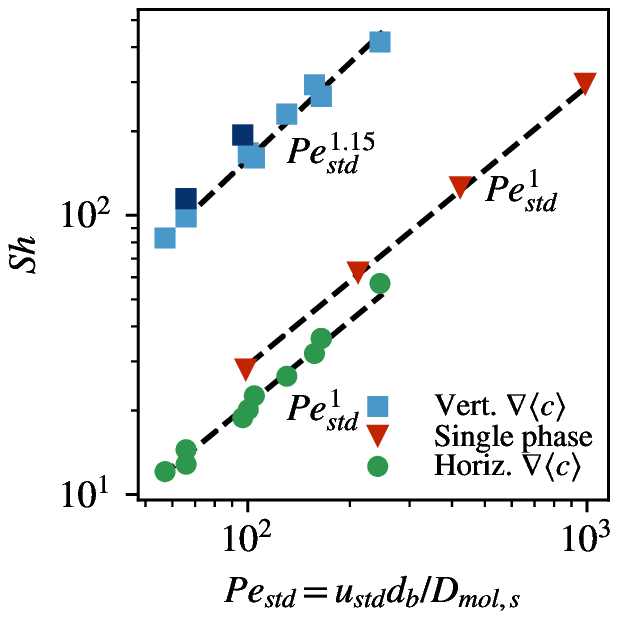}
    \end{overpic}
  \end{subfigure}
  \quad
  \begin{subfigure}[t]{0.33\linewidth}
  \centering
  \caption{}
    \label{fig:NuPe_phi}
    \begin{overpic}[trim = .3cm .3cm 0.3cm 0.3cm,clip,width=\linewidth]{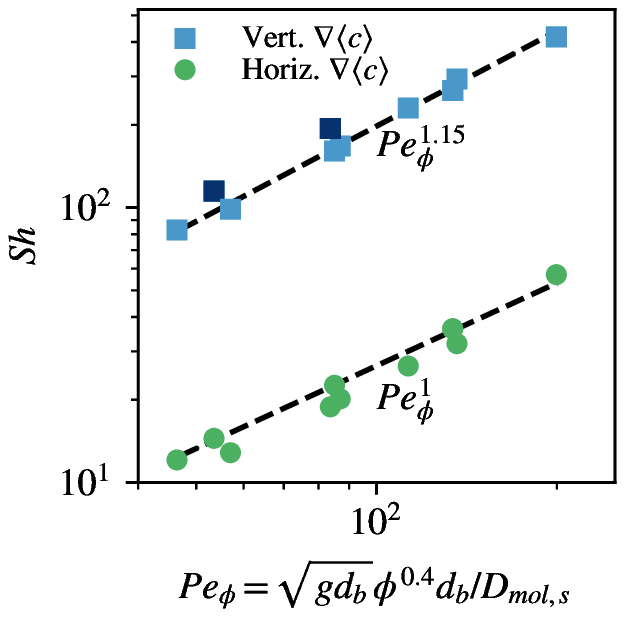}
    \end{overpic}
  \end{subfigure}
  \begin{subfigure}[t]{0.3\linewidth}
  \centering
  \caption{}
    \label{fig:D_conv_phi}
    \begin{overpic}[trim = .3cm .3cm 0.3cm 0.3cm,clip,width=\linewidth]{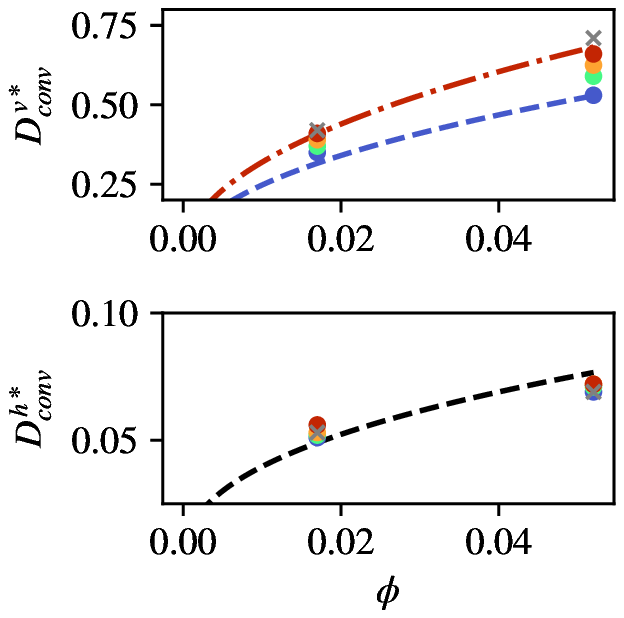}
    \end{overpic}
  \end{subfigure}
    \caption{Convective contribution to the effective diffusivity for all our simulation cases. The darker blue squares in (a) and (b) indicate the cases A5 and B5 with $\nabla^v \langle c \rangle$ and $\mathit{Sc}_g=7$. (a): $\mathit{Sh}$ against $\mathit{Pe}_{\mathit{std}}$ based on the total standard deviation of velocity fluctuations in the liquid. (b): $\mathit{Sh}$ against $\mathit{Pe}_{\phi}$ based on a priori known parameters of the suspension. (c): Nondimensional $\mathsf{D}^*_{\mathit{conv}} = \mathsf{D}_{\mathit{conv}} / (\sqrt{g d_b}d_b)$ against $\phi$ for $\nabla^v \langle c \rangle$ (top panel) and $\nabla^h \langle c \rangle$ (bottom panel). In both panels, the colour of the points represents the same $\mathit{Sc}$-numbers as in \cref{fig:prod}. In the top panel, the proposed scalings for $\mathsf{D}^v_{\mathit{conv}}$ are the dash-dotted line (lowest $D_{\mathit{mol,s}}$, case A4) and the dashed line (highest $D_{\mathit{mol,s}}$, case B1). The dashed line in the bottom panel is the proposed scaling for $\mathsf{D}^h_{\mathit{conv}}$.}
    \label{fig:NuPe}
\end{figure}

We define the Sherwood number as $\mathit{Sh} = \mathsf{D}_{\mathit{conv}} / D_{\mathit{mol,s}}$ that represents the relative importance of the convective contribution to the molecular contribution in the effective diffusivity of the scalar (in heat transfer, the same definition applies to the Nusselt number). In single-phase isotropic turbulence $\mathit{Sh} \propto \mathit{Pe}_{\mathit{std}} = u_{\mathit{std}}l_u/D_{\mathit{mol,s}}$ according to theory and DNS \citep{Gotoh2012} at high Péclet $\mathit{Pe}_{\mathit{std}}$ numbers where diffusion is negligible ($l_u$ is the velocity integral length scale). The same scaling is found at high $\mathit{Pe_d=V_d d_b / D_{\mathit{mol,s}}}$ (for cases with both $\nabla^v \langle c \rangle$ and $\nabla^h \langle c \rangle$) for bubbly flows without fully developed bubble-induced turbulence and with a constant scalar diffusivity \citep{Loisy2018}. \Cref{fig:NuPe_std} shows that indeed $\mathit{Sh} \propto \mathit{Pe}_{\mathit{std}}$ for our single phase and bubbly flow simulations with $\nabla^h \langle c \rangle$. Here, we have used $l_u = d_b$ as the length scale for $\mathit{Pe}_{\mathit{std}}$ in the bubbly flow cases and the forcing length scale in the single-phase simulations.
he higher $\mathit{Sh}$ for cases with $\nabla^v \langle c \rangle$ (compared to cases with $\nabla^h \langle c \rangle$) can be explained by the preferential vertical motion of the bubbles due to buoyancy. The ratio of the liquid velocity components in the vertical and horizontal directions is $u^v_{l,\mathit{std}}/u^h_{\mathit{l,std}}\approx1.4$ (where we obtain for $\phi=1.7\%: u^v_{l,\mathit{std}}=0.29, u^h_{l,\mathit{std}}=0.21$ and for $\phi=5.2\%: u^v_{l,\mathit{std}}=0.42, u^h_{l,\mathit{std}}=0.31$) for all our bubbly flow cases and where the ratios of the vertical to horizontal components are in excellent agreement with experiments \citep{Riboux2010}. Contrarily, in single-phase isotropic turbulence we have $u_{l,\mathit{std}}=u^v_{l,\mathit{std}}=u^h_{l,\mathit{std}}=0.3$. These values suggest that the scalar flux in the single-phase turbulence should be similar to or higher than that in the bubbly flow with $\nabla^h \langle c \rangle$ (where the scalar flux is proportional to $u^h_{l,\mathit{std}}$). The scalar flux in the single-phase turbulence should, however, be less than that in the bubbly flow with $\nabla^v \langle c \rangle$ (where the total scalar flux is governed by $u^v_{l,\mathit{std}}$ and the additional mechanisms acting in the vertical direction discussed in \cref{sec:char_scalar_dyn}).

In the bubbly flow cases with $\nabla^v \langle c \rangle$ we observe approximately $\mathit{Sh} \propto \mathit{Pe}^{1.15}_{\mathit{std}}$ indicating that molecular diffusion is not negligible. Using the scaling $u_{\mathit{std}} \propto V_0\phi^{0.4}$ found in \cite{Risso2002} and \cite{Riboux2010} and noting that the rise velocity of a single bubble $V_0 / (\sqrt{g d_b})=O(1)$ we can define $Pe_{\phi} = \sqrt{g d_b}\phi^{0.4}d_b/D_{\mathit{mol,s}}$ based on a priori known parameters. \Cref{fig:NuPe_phi} shows the $\mathit{Sh}$ against $\mathit{Pe}_\phi$ for our bubbly flow cases where we again observe $\mathit{Sh}^v \propto \mathit{Pe}^{1.15}_{\phi}$ and $\mathit{Sh}^h \propto \mathit{Pe}^{1}_{\phi}$. Solving for $\mathsf{D}_{\mathit{conv}}$ in $\mathit{Sh}$ of the latter scalings gives $\mathsf{D}^v_{\mathit{conv}} \propto (\sqrt{g d_b}d_b)^{1.15}\phi^{0.46}/D_{\mathit{mol,s}}^{0.15}$ and $\mathsf{D}^h_{\mathit{conv}} \propto \sqrt{g d_b}\phi^{0.4}d_b$. These are similar dependencies on $\phi$ as found experimentally in \cite{Almeras2015} for a low-diffusive dye ($\mathsf{D}_{\mathit{conv}}\propto \phi^{0.4}$ at lower $\phi$) and by \cite{Gvozdic2018} for heat transport ($\mathsf{D}_{\mathit{conv}} \propto \phi^{0.45}$ up to $\phi=5\%$).

The $\mathsf{D}_{\mathit{conv}}$ for each simulation case are shown in \cref{fig:D_conv_phi}. In cases A1-A4 and B1-B4 (colored dots), the $\mathit{Sc}_l$-number is increased from $0.7$ to $7$ while the $\mathit{Sc}_g=0.7$ is maintained. For these cases there is a monotonic increase of both $\mathsf{D}^v_{\mathit{conv}}$ and $\mathsf{D}^h_{\mathit{conv}}$ with $\mathit{Sc}_l$. However, in cases A5 and B5 (grey crosses) we increase the $\mathit{Sc}_g=7$ while specifying $\mathit{Sc}_l=0.7$. These parameters result in the highest $\mathsf{D}^v_{\mathit{conv}}$ for the respective cases A1-5 and B1-5 but a $\mathsf{D}^h_{\mathit{conv}}$ closer to the lower $\mathit{Sc}_l$-numbers.

The influence of the $\mathit{Sc}_g$ on $\bm{\mathsf{D}}_{\mathit{conv}}$ can be explained by considering the bubbles as a source of scalar disturbance $c'$. When a bubble moves in the direction of the imposed scalar gradient, the scalar disturbance in the bubble increases proportional to the last term on the r.h.s of \cref{eq:scalar_transport}. The time it takes for the scalar in the bubble to reach an equilibrium with the liquid is proportional to the characteristic equalisation time $t_e = R^2/D_{g}$. A higher $\mathit{Sc}_g$-number thus implies a longer $t_e$ and therefore a higher average value of $c'$ in the gas phase contribution $\phi\langle \boldsymbol{u}'c' \rangle_g$ to $\bm{\mathsf{D}}_{\mathit{conv}}$ in \cref{eq:D_conv}. In our cases, the liquid phase contribution in \cref{eq:D_conv} is almost independent on $\mathit{Sc}_g$. The influence of the $\mathit{Sc}_g$ on $\bm{\mathsf{D}}_{\mathit{conv}}$ is evident in the cases with $\nabla^v \langle c \rangle$ since the bubbles have an average velocity in the vertical direction due to buoyancy. The high average gas phase velocity in the gradient direction induces large average gas phase scalar fluctuations. The high gas phase velocity and scalar fluctuations cause a significant gas phase contribution $\phi\langle \boldsymbol{u}'c' \rangle_g$ to $\mathsf{D}^{v}_{\mathit{conv}}$ although the parameter $\phi$ is small.

The proposed scalings for $\mathsf{D}_{\mathit{conv}}$ are shown in \cref{fig:D_conv_phi} with a proportionality constant of $0.94$ for $\mathsf{D}^v_{\mathit{conv}}$ and $0.25$ for $\mathsf{D}^h_{\mathit{conv}}$. In the top panel, the dash-dotted line represents the scaling for $\mathsf{D}^v_{\mathit{conv}}$ using the lowest $D_{\mathit{mol,s}}$ (case A4) and the dashed line is the same scaling using the highest $D_{\mathit{mol,s}}$ (case B1). All other cases with $\mathit{Sc}_g=0.7$ fall between these limits. Because of the previously discussed effects of $\mathit{Sc}_g$ on $\mathsf{D}^v_{\mathit{conv}}$, the results for $\mathit{Sc}_g=7$ (grey crosses) are not in such good agreement with the proposed scalings. More data is however needed to study the effects of $\mathit{Sc}_g$ on $\mathsf{D}^v_{\mathit{conv}}$. The bottom panel of \cref{fig:D_conv_phi} shows the scaling for $\mathsf{D}^h_{\mathit{conv}}$ and all cases show that the effects of molecular diffusion (the $\mathit{Sc}$-numbers) are not as significant.

\section{Conclusions}\label{Sec:conclusions}

We have performed DNS of bubbly flows with passive scalars and shown a transition of the scalar spectra from a $k^{-5/3}$ scaling (experienced in single-phase isotropic turbulence) to a $k^{-3}$ scaling (in the case of a mean scalar gradient in the vertical direction). In the investigated parameter ranges, we find that the transition length scale is comparable to or below the bubble diameter ($k \geq k_{d_b}$) and decreases with the liquid scalar diffusivity (increasing $\mathit{Sc}_l$). We compute the scalar spectra budget and show that the scalar fluctuations are produced by the mean scalar gradient at the scales above the bubble diameter ($k < k_{d_b}$), while the production term decays as $k^{-3}$ for $k > k_{d_b}$. This is opposed to the velocity fluctuations that are continuously produced and directly dissipated at scales $k > k_{d_b}$. The observed scalar production behaviour is valid in the presence of a mean scalar gradient. However, if the scalar in the liquid is injected/consumed at the bubble surfaces, we expect a scalar production peak at $k_{d_b}$ and hence a production term more similar to the momentum transfer. At length scales below the bubble diameter, the scalar transfer term in the budget equation shows a $k^{-1}$ scaling in a narrow range of wavenumbers inducing the $k^{-3}$ scaling of the scalar spectra for cases with a mean scalar gradient in the vertical direction. These findings are in agreement with the hypothesis proposed in \cite{dung2022emergence} about the physical mechanisms behind the $k^{-3}$ scaling found in that experimental work for heat transport in bubbly flows with an active turbulent grid. We also examine the scalar effective convective diffusivity and the Sherwood number of the bubble suspension and find that $\mathit{Sh} \propto \mathit{Pe}_\phi$ (and consequently $\mathsf{D}^h_{\mathit{conv}} \propto \sqrt{g d_b}\phi^{0.4}d_b$) for an imposed scalar gradient in the horizontal direction. For a vertical scalar gradient, we find however $\mathit{Sh} \propto \mathit{Pe}^{1.15}_\phi$ and $\mathsf{D}^v_{\mathit{conv}} \propto (\sqrt{g d_b}d_b)^{1.15}\phi^{0.46}/D_{\mathit{mol,s}}^{0.15}$ for a constant gas scalar diffusivity. These scalings are based on a priori known parameters and extend the model proposed by \cite{Almeras2015}. We find that the gas scalar diffusivity significantly influences the average gas scalar fluctuations when the scalar gradient is in the vertical direction. This effect modifies $\mathsf{D}^v_{\mathit{conv}}$ and should be considered when developing improved models. Future investigations are needed to investigate how different parameters, like $\mathit{Ga}$ and $\mathit{Eo}$ numbers, volume fraction, molecular diffusivities of the two phases and different scalar injection mechanisms influence the observed scalar dynamics.

\paragraph{Funding.}{The authors acknowledge support by the Swedish Research Council (Vetenskapsr\aa det), grant VR 2017-05031. The authors gratefully acknowledge the HPC RIVR consortium (www.hpc-rivr.si) and EuroHPC JU (eurohpc-ju.europa.eu) for funding this research by providing computing resources of the HPC system Vega at the Institute of Information Science (www.izum.si). Other computational resources have been provided by the Swedish National Infrastructure for Computing (SNIC) at NSC partially funded by the Swedish Research Council through grant agreement no. 2018-05973.}

\paragraph{Declaration of interests.}{The authors report no conflict of interest.}

\paragraph{Author ORCID.}{N. Hidman, https://orcid.org/0000-0001-9973-9451; H. Str\"om, https://orcid.org/0000-0002-8581-5174; S. Sasic, https://orcid.org/0000-0001-6383-4772; G. Sardina, https://orcid.org/0000-0002-9172-6311}


\bibliographystyle{jfm}
\bibliography{mybibfile}

\begin{thebibliography}{36}
\expandafter\ifx\csname natexlab\endcsname\relax\def\natexlab#1{#1}\fi
\def\au#1{#1} \def\ed#1{#1} \def\yr#1{#1}\def\at#1{#1}\def\jt#1{\textit{#1}}
  \def\bt#1{#1}\def\bvol#1{\textbf{#1}} \def\vol#1{#1} \def\pg#1{#1}
  \def\publ#1{#1}\def\arxiv#1{#1}\def\org#1{#1}\def\st#1{\textit{#1}}

\bibitem[Alm{\'e}ras {\em et~al.\/}(2016)Alm{\'e}ras, Cazin, Roig, Risso,
  Augier \& Plais]{Almeras2016}
{\sc \au{Alm{\'e}ras, E.}, \au{Cazin, S.}, \au{Roig, V.}, \au{Risso, F.},
  \au{Augier, F.} \& \au{Plais, C.}} \yr{2016}  \at{Time-resolved measurement
  of concentration fluctuations in a confined bubbly flow by lif}.
  \jt{International Journal of Multiphase Flow}  \bvol{83},  \pg{153--161}.

\bibitem[Alm{\'e}ras {\em et~al.\/}(2015)Alm{\'e}ras, Risso, Roig, Cazin, Plais
  \& Augier]{Almeras2015}
{\sc \au{Alm{\'e}ras, E.}, \au{Risso, F.}, \au{Roig, V.}, \au{Cazin, S.},
  \au{Plais, C.} \& \au{Augier, F.}} \yr{2015}  \at{Mixing by bubble-induced
  turbulence}.  \jt{Journal of Fluid Mechanics}  \bvol{776},  \pg{458--474}.

\bibitem[Alm{\'e}ras {\em et~al.\/}(2018)Alm{\'e}ras, Risso, Roig, Plais \&
  Augier]{Almeras2018}
{\sc \au{Alm{\'e}ras, E.}, \au{Risso, F.}, \au{Roig, V.}, \au{Plais, C.} \&
  \au{Augier, F.}} \yr{2018}  \at{Mixing mechanism in a two-dimensional bubble
  column}.  \jt{Physical Review Fluids}  \bvol{3}~(7),  \pg{074307}.

\bibitem[Batchelor(1959)]{Batchelor1959}
{\sc \au{Batchelor, G.K.}} \yr{1959}  \at{Small-scale variation of convected
  quantities like temperature in turbulent fluid part 1. general discussion and
  the case of small conductivity}.  \jt{Journal of Fluid Mechanics}
  \bvol{5}~(1),  \pg{113–133}.

\bibitem[Bunner \& Tryggvason(2002)]{Bunner2002}
{\sc \au{Bunner, B.} \& \au{Tryggvason, G.}} \yr{2002}  \at{Dynamics of
  homogeneous bubbly flows part 1. rise velocity and microstructure of the
  bubbles}.  \jt{Journal of Fluid Mechanics}  \bvol{466},  \pg{17--52}.

\bibitem[Corrsin(1951)]{Corrsin1951}
{\sc \au{Corrsin, S.}} \yr{1951}  \at{On the spectrum of isotropic temperature
  fluctuations in an isotropic turbulence}.  \jt{Journal of Applied Physics}
  \bvol{22}~(4),  \pg{469--473}.

\bibitem[Dung {\em et~al.\/}(2023)Dung, Waasdorp, Sun, Lohse \&
  Huisman]{dung2022emergence}
{\sc \au{Dung, O.Y.}, \au{Waasdorp, P.}, \au{Sun, C.}, \au{Lohse, D.} \&
  \au{Huisman, S.G.}} \yr{2023}  \at{The emergence of bubble-induced scaling in
  thermal spectra in turbulence}.  \jt{Journal of Fluid Mechanics}  \bvol{958},
   \pg{A5}.

\bibitem[Gotoh \& Watanabe(2012)]{Gotoh2012}
{\sc \au{Gotoh, T.} \& \au{Watanabe, T.}} \yr{2012}  \at{Scalar flux in a
  uniform mean scalar gradient in homogeneous isotropic steady turbulence}.
  \jt{Physica D: Nonlinear Phenomena}  \bvol{241}~(3),  \pg{141--148}.

\bibitem[Gvozdi{\'c} {\em et~al.\/}(2018)Gvozdi{\'c}, Alm{\'e}ras, Mathai, Zhu,
  van Gils, Verzicco, Huisman, Sun \& Lohse]{Gvozdic2018}
{\sc \au{Gvozdi{\'c}, B.}, \au{Alm{\'e}ras, E.}, \au{Mathai, V.}, \au{Zhu, X.},
  \au{van Gils, D.P.}, \au{Verzicco, R.}, \au{Huisman, S.G.}, \au{Sun, C.} \&
  \au{Lohse, D.}} \yr{2018}  \at{Experimental investigation of heat transport
  in homogeneous bubbly flow}.  \jt{Journal of Fluid Mechanics}  \bvol{845},
  \pg{226--244}.

\bibitem[Hidman {\em et~al.\/}(2022)Hidman, Str{\"o}m, Sasic \&
  Sardina]{hidman2022}
{\sc \au{Hidman, N.}, \au{Str{\"o}m, H.}, \au{Sasic, S.} \& \au{Sardina, G.}}
  \yr{2022}  \at{The lift force on deformable and freely moving bubbles in
  linear shear flows}.  \jt{Journal of Fluid Mechanics}  \bvol{952},  \pg{A34}.

\bibitem[Innocenti {\em et~al.\/}(2021)Innocenti, Jaccod, Popinet \&
  Chibbaro]{Innocenti2021}
{\sc \au{Innocenti, A.}, \au{Jaccod, A.}, \au{Popinet, S.} \& \au{Chibbaro,
  S.}} \yr{2021}  \at{Direct numerical simulation of bubble-induced
  turbulence}.  \jt{Journal of Fluid Mechanics}  \bvol{918},  \pg{A23}.

\bibitem[Lance \& Bataille(1991)]{lance_bataille_1991}
{\sc \au{Lance, M.} \& \au{Bataille, J.}} \yr{1991}  \at{Turbulence in the
  liquid phase of a uniform bubbly air–water flow}.  \jt{Journal of Fluid
  Mechanics}  \bvol{222},  \pg{95–118}.

\bibitem[Loisy(2016)]{Loisy_thesis}
{\sc \au{Loisy, A.}} \yr{2016}  \at{{Direct numerical simulation of bubbly
  flows: coupling with scalar transport and turbulence}}. Theses,
  {Universit{\'e} de Lyon}.

\bibitem[Loisy {\em et~al.\/}(2017)Loisy, Naso \& Spelt]{Loisy2017}
{\sc \au{Loisy, A.}, \au{Naso, A.} \& \au{Spelt, P.D.}} \yr{2017}
  \at{Buoyancy-driven bubbly flows: ordered and free rise at small and
  intermediate volume fraction}.  \jt{Journal of Fluid Mechanics}  \bvol{816},
  \pg{94--141}.

\bibitem[Loisy {\em et~al.\/}(2018)Loisy, Naso \& Spelt]{Loisy2018}
{\sc \au{Loisy, A.}, \au{Naso, A.} \& \au{Spelt, P.D.}} \yr{2018}  \at{The
  effective diffusivity of ordered and freely evolving bubbly suspensions}.
  \jt{Journal of Fluid Mechanics}  \bvol{840},  \pg{215--237}.

\bibitem[Lumley(1964)]{Lumley1964}
{\sc \au{Lumley, J.L.}} \yr{1964}  \at{The spectrum of nearly inertial
  turbulence in a stably stratified fluid}.  \jt{Journal of the Atmospheric
  Sciences}  \bvol{21}~(1),  \pg{99--102}.

\bibitem[Mendez-Diaz {\em et~al.\/}(2013)Mendez-Diaz, Serrano-Garcia, Zenit \&
  Hernandez-Cordero]{Mendez-Diaz2013}
{\sc \au{Mendez-Diaz, S.}, \au{Serrano-Garcia, J.}, \au{Zenit, R.} \&
  \au{Hernandez-Cordero, J.}} \yr{2013}  \at{Power spectral distributions of
  pseudo-turbulent bubbly flows}.  \jt{Physics of Fluids}  \bvol{25}~(4),
  \pg{043303}.

\bibitem[Mercado {\em et~al.\/}(2010)Mercado, Gomez, Van~Gils, Sun \&
  Lohse]{mercado2010}
{\sc \au{Mercado, J.M.}, \au{Gomez, D.C.}, \au{Van~Gils, D.}, \au{Sun, C.} \&
  \au{Lohse, D.}} \yr{2010}  \at{On bubble clustering and energy spectra in
  pseudo-turbulence}.  \jt{Journal of Fluid Mechanics}  \bvol{650},
  \pg{287--306}.

\bibitem[Monin \& Yaglom(1975)]{monin1975}
{\sc \au{Monin, A.} \& \au{Yaglom, A.}} \yr{1975} {\em Statistical Fluid
  Mechanics\/}.  \publ{MIT Press, Cambridge, MA}.

\bibitem[Mudde(2005)]{Mudde2005}
{\sc \au{Mudde, R.F.}} \yr{2005}  \at{Gravity-driven bubbly flows}.  \jt{Annu.
  Rev. Fluid Mech.}  \bvol{37},  \pg{393--423}.

\bibitem[Pandey {\em et~al.\/}(2020)Pandey, Ramadugu \& Perlekar]{Pandey2020}
{\sc \au{Pandey, V.}, \au{Ramadugu, R.} \& \au{Perlekar, P.}} \yr{2020}
  \at{Liquid velocity fluctuations and energy spectra in three-dimensional
  buoyancy-driven bubbly flows}.  \jt{Journal of Fluid Mechanics}  \bvol{884},
  \pg{R6}.

\bibitem[Pope(2001)]{pope2001}
{\sc \au{Pope, S.B.}} \yr{2001} {\em Turbulent flows\/}.  \publ{IOP
  Publishing}.

\bibitem[Popinet(2009)]{POPINET2009}
{\sc \au{Popinet, S.}} \yr{2009}  \at{An accurate adaptive solver for
  surface-tension-driven interfacial flows}.  \jt{Journal of Computational
  Physics}  \bvol{228}~(16),  \pg{5838 -- 5866}.

\bibitem[Popinet(2015)]{Popinet2015}
{\sc \au{Popinet, S.}} \yr{2015}  \at{A quadtree-adaptive multigrid solver for
  the serre-green-naghdi equations}.  \jt{Journal of Computational Physics}
  \bvol{302},  \pg{336--358}.

\bibitem[Popinet(2018)]{Popinet2018}
{\sc \au{Popinet, S.}} \yr{2018}  \at{Numerical models of surface tension}.
  \jt{Annual Review of Fluid Mechanics}  \bvol{50},  \pg{49--75}.

\bibitem[Prakash {\em et~al.\/}(2016)Prakash, Mercado, van Wijngaarden,
  Mancilla, Tagawa, Lohse \& Sun]{prakash2016}
{\sc \au{Prakash, V.N.}, \au{Mercado, J.M.}, \au{van Wijngaarden, L.},
  \au{Mancilla, E.}, \au{Tagawa, Y.}, \au{Lohse, D.} \& \au{Sun, C.}} \yr{2016}
   \at{Energy spectra in turbulent bubbly flows}.  \jt{Journal of Fluid
  Mechanics}  \bvol{791},  \pg{174--190}.

\bibitem[Rensen {\em et~al.\/}(2005)Rensen, Luther \& Lohse]{rensen2005}
{\sc \au{Rensen, J.}, \au{Luther, S.} \& \au{Lohse, D.}} \yr{2005}  \at{The
  effect of bubbles on developed turbulence}.  \jt{Journal of Fluid Mechanics}
  \bvol{538},  \pg{153--187}.

\bibitem[Riboux {\em et~al.\/}(2010)Riboux, Risso \& Legendre]{Riboux2010}
{\sc \au{Riboux, G.}, \au{Risso, F.} \& \au{Legendre, D.}} \yr{2010}
  \at{Experimental characterization of the agitation generated by bubbles
  rising at high reynolds number}.  \jt{Journal of Fluid Mechanics}
  \bvol{643},  \pg{509--539}.

\bibitem[Risso(2018)]{Risso2018}
{\sc \au{Risso, F.}} \yr{2018}  \at{Agitation, mixing, and transfers induced by
  bubbles}.  \jt{Annual Review of Fluid Mechanics}  \bvol{50},  \pg{25--48}.

\bibitem[Risso \& Ellingsen(2002)]{Risso2002}
{\sc \au{Risso, F.} \& \au{Ellingsen, K.}} \yr{2002}  \at{Velocity fluctuations
  in a homogeneous dilute dispersion of high-reynolds-number rising bubbles}.
  \jt{Journal of Fluid Mechanics}  \bvol{453},  \pg{395--410}.

\bibitem[Rycroft(2009)]{Rycroft2009}
{\sc \au{Rycroft, C.H.}} \yr{2009}  \at{Voro++: A three-dimensional voronoi
  cell library in c++}.  \jt{Chaos: An Interdisciplinary Journal of Nonlinear
  Science}  \bvol{19}~(4),  \pg{041111}.

\bibitem[Sardina {\em et~al.\/}(2015)Sardina, Picano, Brandt \&
  Caballero]{Sardina2015}
{\sc \au{Sardina, G.}, \au{Picano, F.}, \au{Brandt, L.} \& \au{Caballero, R.}}
  \yr{2015}  \at{Continuous growth of droplet size variance due to condensation
  in turbulent clouds}.  \jt{Phys. Rev. Lett.}  \bvol{115},  \pg{184501}.

\bibitem[Scardovelli \& Zaleski(1999)]{Scardovelli1999}
{\sc \au{Scardovelli, R.} \& \au{Zaleski, S.}} \yr{1999}  \at{Direct numerical
  simulation of free-surface and interfacial flow}.  \jt{Annual Review of Fluid
  Mechanics}  \bvol{31}~(1),  \pg{567--603}.

\bibitem[Tagawa {\em et~al.\/}(2013)Tagawa, Roghair, Prakash, van
  Sint~Annaland, Kuipers, Sun \& Lohse]{tagawa2013}
{\sc \au{Tagawa, Y.}, \au{Roghair, I.}, \au{Prakash, V.N.}, \au{van
  Sint~Annaland, M.}, \au{Kuipers, H.}, \au{Sun, C.} \& \au{Lohse, D.}}
  \yr{2013}  \at{The clustering morphology of freely rising deformable
  bubbles}.  \jt{Journal of Fluid Mechanics}  \bvol{721},  \pg{R2}.

\bibitem[Talley {\em et~al.\/}(2017)Talley, Zimmer \& Bolotnov]{Talley2017}
{\sc \au{Talley, M.L.}, \au{Zimmer, M.D.} \& \au{Bolotnov, I.A.}} \yr{2017}
  \at{Coalescence prevention algorithm for level set method}.  \jt{Journal of
  Fluids Engineering}  \bvol{139}~(8).

\bibitem[Tryggvason {\em et~al.\/}(2011)Tryggvason, Scardovelli \&
  Zaleski]{Tryggvason2011}
{\sc \au{Tryggvason, G.}, \au{Scardovelli, R.} \& \au{Zaleski, S.}} \yr{2011}
  {\em Direct numerical simulations of gas--liquid multiphase flows\/}.
  \publ{Cambridge University Press}.

\end{thebibliography}

\end{document}